\documentclass[fleqn,usenatbib]{mnras}

\usepackage[T1]{fontenc}

\DeclareRobustCommand{\DA}[3]{#2}
\let\DAthebibliography\thebibliography
\def\thebibliography{\DeclareRobustCommand{\DA}[3]{##3}\DAthebibliography}

\usepackage{graphicx}	
\usepackage{amsmath}	
\usepackage{amssymb}	
\usepackage{float}
\usepackage{caption}
\usepackage{subcaption}
\usepackage{soul}
\usepackage{lineno}

\usepackage{newtxtext,newtxmath}
\usepackage{xcolor}

\newcommand{\magphys}{\textsc{Magphys}}

\title[Energy balance with UV-FIR offsets]{Energy balance SED modelling can be effective at high redshifts regardless of UV-FIR offsets}

\author[P.~Haskell et al.]{
P.~Haskell$^{1}$\thanks{E-mail: ph18aai@herts.ac.uk},
D.~J.~B.~Smith$^{1},$
R.~K.~Cochrane$^{2},$
C.~C.~Hayward$^{2}$
and D.~Angl\'es-Alc\'azar$^3{}^,{}^2$
\\
$^{1}$Centre for Astrophysics Research, University of Hertfordshire, College Lane, Hatfield AL10 9AB, UK
\\
$^{2}$Center for Computational Astrophysics, Flatiron Institute, 162 Fifth Avenue, New York, NY 10010, USA
\\
$^{3}$Department of Physics, University of Connecticut, 196 Auditorium Road, U-3046, Storrs, CT 06269-3046, USA 
}

\date{Accepted XXX. Received YYY; in original form ZZZ}

\pubyear{2023}

\begin{document}
\label{firstpage}
\pagerange{\pageref{firstpage}--\pageref{lastpage}}
\maketitle

\newcommand{\comment}[1]{}

\begin{abstract}
{\color{black}Recent works have suggested that energy balance spectral energy distribution (SED) fitting codes may be of limited use for studying high-redshift galaxies for which the observed ultraviolet and far-infrared emission are offset (spatially `decoupled'). It has been proposed that such offsets could lead energy balance codes to miscalculate the overall energetics, preventing them from recovering such galaxies' true properties. 
In this work, we test how well the SED fitting code \magphys\ can recover the stellar mass, star formation rate (SFR), specific SFR, dust mass and luminosity by fitting 6,706 synthetic SEDs generated from four zoom-in simulations of dusty, high-redshift galaxies from the FIRE project via dust continuum radiative transfer. 
Comparing our panchromatic results (using wavelengths 0.4--500\,$\mu$m, and spanning $1<z<8$) with fits based on either the starlight ($\lambda_\mathrm{eff} \le 2.2\,\mu$m) or dust ($\ge 100\,\mu$m) alone, we highlight the power of considering the full range of multi-wavelength data alongside an energy balance criterion. Overall, we obtain  acceptable fits for 83 per cent of the synthetic SEDs, though the success rate falls rapidly beyond $z \approx 4$, in part due to the sparser sampling of the priors at earlier times since SFHs must be physically plausible (i.e. shorter than the age of the Universe). We use the ground truth from the simulations to show that when the quality of fit is acceptable, the fidelity of \magphys\ estimates is independent of the degree of UV\slash FIR offset, with performance very similar to that previously reported for local galaxies.}
\end{abstract}

\begin{keywords}
galaxies: high-redshift - galaxies: fundamental parameters - methods: data analysis
\end{keywords}




\section{Introduction}

Spectral energy distribution (SED) fitting offers a powerful method of estimating galaxy physical properties from photometry. 
{\color{black}SED fitting programs take as input the available photometry, which can be $>30$ bands in the best studied fields to $<10$ elsewhere, then use models of varying complexity to infer the shape of the full SED and hence the underlying physical properties \citep[for an introduction to SED fitting see e.g.][]{Walcher2011,Conroy2013}}.  

The energy balance code \magphys\ (\citealt{daCunha2008} - hereafter DC08) 
performs $\chi^2$ fitting using two sets of pre-built libraries of model SEDs with a representative range of SFHs and dust models for star-forming galaxies. 
The energy balance criterion works in such a way that \magphys\ considers only combinations of SFH and dust emission that are energetically consistent, in the sense that the energy absorbed by dust in the rest-frame UV is re-radiated in the FIR. During the fit, \magphys\ finds the SFH and dust model that best fits the data, and calculates probability density functions (PDFs) for a variety of property values by marginalising over all of the models which satisfy the energy balance criterion.  
\par 
To determine the fidelity of the properties derived from SED fitting, three testing techniques have been used in previous studies. The first is to compare the derived physical parameters to those derived using simpler methods. DC08 tested how well \magphys\ could fit observations from the \textit{Spitzer} Infrared Nearby Galaxy Survey \citep[SINGS][]{sings}, producing acceptable best-fit $\chi^2$ results for 63 of the 66 galaxies. They also tested how well \magphys\ could recover the properties of 100 of its own, randomly selected, models with noise added to the photometry. Here, $M_{\mathrm{star}}$, SFR and $L_{{\mathrm{dust}}}$ were reported to be recovered to a high degree of accuracy. Similarly, \citet{Noll2009} tested the alternative energy balance SED fitting code {\sc{CIGALE}} {\color{black}\citep{Boquien2019}} using the SINGS galaxies, replacing DC08’s UBV observations with those from \citet{Munoz-Mateos2009}. Here, $L_{{\mathrm{dust}}}$ estimates compared well ($\pm 0.03$ dex) with those derived by \citet{Draine2007}, similarly the SFR estimates compared well ($0.06 \pm 0.05 $ dex) with those provided by \citet[]{Kennicutt1998} based on H$\alpha$ emission (e.g. \citealt{Kennicutt1998a}). 
\par 
An alternative testing technique is to compare the results of different fitting programs when applied to the same dataset. This will not provide evidence that the results are correct, but does give confidence that a given code performs similarly to others. Best et al. (2022 - in preparation) tested three energy balance based fitters - \magphys, {\sc{CIGALE}} and {\sc{BAGPIPES}} \citep{Carnall2018} - together with {\sc{AGNfitter}} \citep{CalistroRivera2016}. The four codes were each used to estimate $M_{\mathrm{star}}$ and SFR for galaxies in the Bo\"{o}tes, Lockman Hole and ELAIS-N1 fields of the LOFAR Two Metre Sky Survey \citep[LoTSS][]{Shimwell2017} deep fields first data release (\citealt{Duncan2021}, \citealt{Kondapally2021}, \citealt{Sabater2021} and \citealt{Tasse2021}). The results of the runs were compared to determine how well they agreed with each other. For galaxies with no AGN,  \magphys, {\sc{CIGALE}} and {\sc{BAGPIPES}} typically agreed to within 0.1 dex for stellar mass, with {\sc{AGNfitter}} differing by 0.3 dex. Similar levels of agreement were found for the SFRs of galaxies found not to contain an AGN. For galaxies with an AGN the situation was more mixed as neither \magphys\ nor {\sc{BAGPIPES}} are designed to handle AGN emission.

\citet{Hunt2019} compared the results of applying \magphys, {\sc{CIGALE}} and {\sc{Grasil}} \citep{Silva1998} to a sample of 61 galaxies from the Key Insights on Nearby Galaxies: a Far-Infrared Survey with \textit{Herschel} (KINGFISH) survey \citep{Kennicutt2011}, including 57 of the SINGS galaxies. They found that stellar masses estimated using 3.6$\mu$m luminosity agreed with all three codes to within 0.2 dex. Similarly, SED derived SFR estimates were within 0.2 dex of those derived using FUV+TIR luminosities and $H\alpha+24\mu m$ luminosities. The results for $M_{\mathrm{dust}}$ were more mixed, with {\sc{Grasil}} giving values 0.3 dex higher than \magphys\ or {\sc{CIGALE}} or the value determined using a single temperature modified black body. A similar approach with an even broader selection of fourteen SED fitting codes was taken by \citet{Pacifici2023}, who found agreement on stellar mass estimates across the ensemble, but some discrepancies in their SFR and dust attenuation results. {\color{black} More recently, \citet{Cheng2023} used a modified version of \magphys\  (\magphys+photo-z; \citealt{Battisti2019}) to determine the photometric redshifts of 16 sub-millimetre galaxies (SMGs). The results were compared to the redshifts derived using EAZY \citep{Brammer2008}, finding that for most sources the results were consistent.}
\par 
The final, and perhaps most promising technique for validating SED fitting is to use simulated galaxies where the ‘right’ answer is known in advance. \citet{Wuyts2009} used the HYPERZ \citep{Bolzonella2000} SED fitting code on GADGET-2 \citep{Springel2005} simulations to recover mass, age, E(B-V) and $A_V$ under a variety of conditions. They concluded that recovery of properties for ellipticals was generally good (residuals between 0.02 and 0.03 dex) with slightly poorer results for disks (residuals of 0.03 to 0.35 dex), with residuals increasing further to 0.02 to 0.54 dex during periods of merger-triggered star formation. \citet[hereafter HS15]{Hayward2015} used \magphys\ on two GADGET-3 \citep{Springel2005} simulations of an isolated disk and a major merger of two disk galaxies at $z = 0.1$. Snapshots were taken at 10$\,$Myr intervals and the radiative transfer code SUNRISE \citep{Jonsson2006} used to produce observations from 7 different lines of sight around the simulation. In both scenarios, the attenuated SED was recovered with an acceptable fit ($\chi^2$ within the 99 per cent confidence threshold; see \citealt{Smith2012b} for details) except for the time around the peak starburst/coalescence phase of the merger simulation. In both scenarios, $L_{{\mathrm{dust}}}$ was recovered well with $M_{{\mathrm{star}}}$ recovered to within 0.3 dex and SFR within 0.2 dex. $M_{{\mathrm{dust}}}$ was recovered less well, but still within 0.3 dex for the isolated galaxy and 0.5 dex for the merger. The conclusion from this study is that these properties of local galaxies can typically be recovered to within a factor of 1.5 – 3. \citet{Smith2018} studied a resolved simulated isolated disk, using spatial resolution as fine as $0.2\,\rm{kpc}$. They found that \magphys\ produced statistically acceptable results for $M_{{\mathrm{star}}}$, $L_{{\mathrm{dust}}}$, SFR, sSFR and $A_V$ for over 99 per cent of pixels within the r-band effective radius. At higher redshifts, \citet[hereafter D20]{Dudzeviciute2020}, used EAGLE (\citealt{Schaye2015}, \citealt{Crain2015}) simulations with {\sc{SKIRT}} generated  photometry (\citealt{Baes2011}, \citealt{Camps2020}) to validate the performance of \magphys\ for studying galaxies with redshifts up to 3.4. They found that \magphys\ gave a remarkably linear correlation with the true (simulated) values, though with significant scatter (at the level of 10, 15 and 30\,per cent for the dust mass, SFR and stellar masses, respectively) and significant systematic offsets (of up to $0.46\pm0.10$\,dex for the recovered stellar mass).
\par 
These studies all provide evidence that SED fitting, particularly energy balance SED fitting, is working remarkably well and providing results often consistent with the ground truth once the uncertainties are accounted for.

{\color{black} 
However, several authors have questioned whether using an energy balance criterion is appropriate when viewing galaxies for which the UV and FIR are spatially offset from one another \citep[e.g.][]{casey2017,Miettinen2017,Simpson2017,Buat2019}. In such cases, while `energy balance' is still expected overall (i.e. energy conservation is presumably not violated), significant spatial decoupling may lead to difficulties in recovering the true properties. Under such circumstances, the attenuation -- and thus the intrinsic UV luminosity -- may be underestimated because the UV-bright, relatively dust-free regions can result in a blue UV-optical slope even if the bulk of the young stars are heavily dust-obscured. 

This concern has recently become testable with the  sub-arcsecond resolution provided by the Atacama Large Millimetre/submillimetre Array (ALMA)\footnote{http://www.alma.info}, enabling direct observation of UV/optical and FIR offsets. There are now numerous papers reporting spatial offsets. \citet{Hodge2016}, \citet{Rujopakarn2016}, \citet{Gomez2018} and \citet{Rujopakarn2019} have discovered kpc offsets between star forming regions and centres of stellar mass while investigating the star formation and dust distributions in $2<z<4.5$ galaxies. Along these lines, \citet{Chen2017} found a significant offset in ALESS67.1, a SMG at $z=2.12$, \citet{Cochrane2021} reported the same in the massive star-forming galaxy SHiZELS-14 at $z=2.24$, and \citet{Bowler2018} detected a 3 kpc offset between the rest-frame FIR and UV emission in the Lyman-break galaxy ID65666 at $z\approx7$.

The concern over the impact of decoupling between the dust and starlight is such that new SED fitting codes such as \texttt{MICHI2} \citep{michi2_coderef} and \texttt{Stardust} \citep{Kokorev2021} mention the \textit{absence} of energy balance as a key advantage in favour of using these codes for studying galaxies where spatial offsets are likely to be a factor. In \citet{Liu2021}, \texttt{MICHI2} produced results very similar to \magphys\ and CIGALE for a sample of high redshift galaxies, with stellar mass and dust luminosity estimates obtained to within 0.2\,-\,0.3\,dex of those obtained using the two energy-balance codes. Similarly, \citet{Kokorev2021} used \texttt{Stardust} to fit 5,000 IR bright galaxies in the GOODS-N and COSMOS fields, producing results which compared well with those derived using CIGALE with a mean $M_\mathrm{dust}$ residual of 0.09\,dex, a mean $L_\mathrm{IR}$ residual of 0.2\,dex and a mean $M_\mathrm{star}$ residual of 0.1\,dex (albeit with a significant scatter of 0.3\,dex). 

An additional test of the likely impact of spatial offsets was conducted by \citet{Seille2022}, who used the CIGALE code to model the Antennae Galaxy, Arp244, which is known to have very different UV and IR distributions \citep{Zhang2010}. \citet{Seille2022} found that the total stellar mass and SFR were consistent, whether they attempted to fit the integrated photometry of the galaxy or sum the results of fitting 58 different regions of Arp244 independently and summed the results (i.e. performance very similar to that found by \citealt{Smith2018} for simulated galaxies without spatial offsets).

In this context, we now seek to further test the efficacy of energy balance SED fitting for these more challenging dusty, high redshift, star-forming galaxies by using high-resolution simulations with differing degrees of spatial offset between the apparent UV/FIR emission.}

This paper is structured as follows. Section 2 describes the tools and methods used to create the observations and to fit the SEDs; Section 3 presents the results of the fitting including the derived values for several galaxy properties; Section 4 discusses these in the context of previous papers and Section 5 summarises the conclusions.
Throughout this work we adopt a standard cosmology with $H_0=70\,$km\,s$^{-1}$\,Mpc$^{-1}$, $\Omega_M=0.3$, and $\Omega_\Lambda=0.7$.


\section{Method}\label{Method}
This section describes the simulation data and the creation of the synthetic observations. It also provides a brief introduction to \magphys, details of the simulations, and how they were subsequently analysed.

\subsection{Computing the SEDs of simulated galaxies}
We analyze a set of 4 cosmological zoom-in simulations from the FIRE project\footnote{\url{http://fire.northwestern.edu}} that were run using the FIRE-2 code \citep{Hopkins2018} down to $z = 1$. The simulations use the code GIZMO \citep{Hopkins2015}\footnote{\url{http://www.tapir.caltech.edu/~phopkins/Site/GIZMO.html}}, with hydrodynamics solved using the mesh-free Lagrangian Godunov ``MFM'' method. Both hydrodynamic and gravitational (force-softening) spatial resolution are set in a fully-adaptive Lagrangian manner with fixed mass resolution. The simulations include cooling and heating from a meta-galactic background and local stellar sources from $T\approx10-10^{10}\,$K; star formation in locally self-gravitating, dense, self-shielding molecular, Jeans-unstable gas; and stellar feedback from OB \&\ AGB mass-loss, SNe Ia \&\ II, multi-wavelength photo-heating and radiation pressure with inputs taken directly from stellar evolution models. The FIRE-2 physics, source code, and all numerical parameters are {\em exactly} identical to those in \citet{Hopkins2018}.

The specific sample of simulations studied in this paper include the halos first presented in \citet{Feldmann2016}.
The FIRE-2 simulations for these halos were introduced, along with a novel on-the-fly treatment of black hole seeding and growth in \citet{Angles2017}. These halos were chosen because
they are representative of the high-redshift, massive, dusty star-forming galaxies found in infrared-selected
observational samples, {\color{black}\citet{Cochrane2019} showing that they present a clumpy dust distribution together with very different morphologies for stellar mass, dust, gas and young stars.}
{\color{black} At $z=2$, the galaxies central to the halos have half-light radii of 0.73, 0.98, 0.81 and 0.91 kpc; for additional information on these galaxies see \citet{Angles2017} as well as \citet{Cochrane2019}, \citet{Wellons2020}, \citet{Parsotan2021} and \citet{Cochrane2022}.}
\par
To generate synthetic SEDs, Monte Carlo dust radiative transfer was performed on each time snapshot of the simulated
galaxies in post-processing using the code SKIRT\footnote{\url{http://www.skirt.ugent.be/}}. SKIRT assigns
single-age stellar population SEDs to star particles in the simulations according to their ages and metallicites.
It then propagates photon packets through the simulated galaxies' ISM to compute the effects of dust absorption,
scattering, and re-emission. {\color{black}Snapshots of the galaxies' evolution were taken at 15 - 25 Myr intervals with each galaxy `observed' from 7 positions that uniformly
sampled inclination angles from view 0 (aligned with the angular momentum vector) in steps of $30^\circ$ to view 6 (anti-aligned).} For full details
of the SKIRT calculations, see \citet[][]{Cochrane2019,Cochrane2022}. This procedure yielded 6,706 SEDs {\color{black} across the four simulated galaxies, spanning $1 < z < 8$.  }
\par
\begin{figure}
\centering
\includegraphics[width=0.47\textwidth]{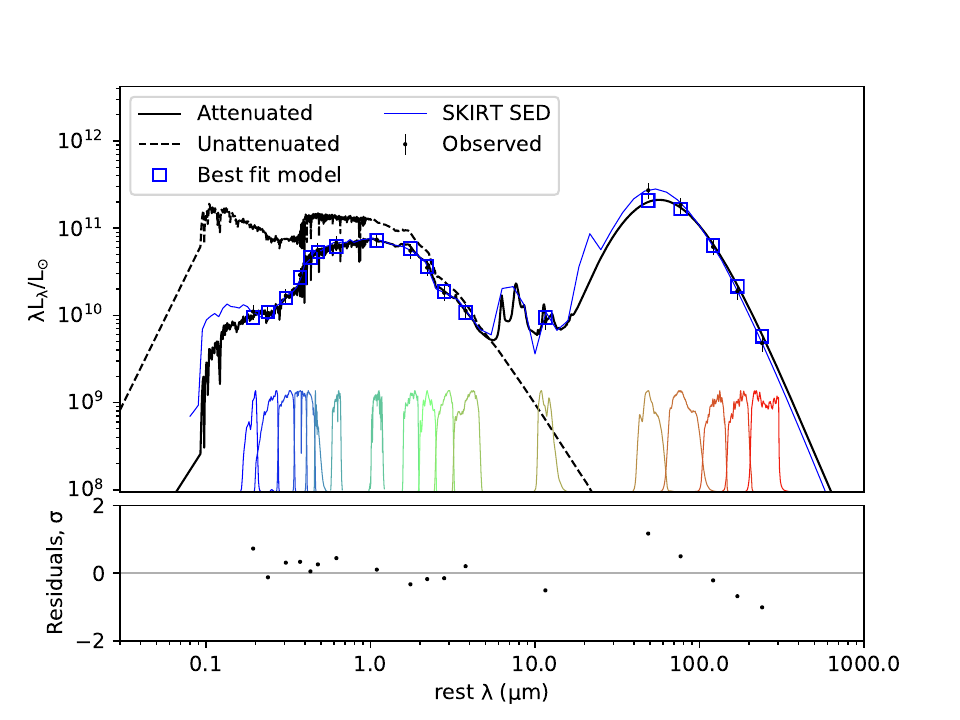}
\caption{An example SED obtained using \magphys, demonstrating the generally close agreement between the true and \magphys-derived SEDs. In the upper panel, the solid black line shows the best-fit \magphys-derived SED, while the dashed black line indicates the \magphys\ estimate of the unattenuated SED; the solid blue line represents the attenuated SED generated by SKIRT. The square markers represent the best-fit photometry, with the SKIRT photometry shown as the points with error bars (as described in the legend). The coloured lines above the lower horizontal axis show the normalised filter curves used in this study. The lower panel shows the resifdual value in $\sigma$ units between each observation and the best-fit SED. The residual value is calculated as (observed flux - model flux)/observed error. This SED corresponds to simulated galaxy A1, snapshot 276, view 0, z=1.00.}
\label{fig:SED_example}
\end{figure}

To compute photometry from the SEDs, we convolved the SEDs with appropriate filter response curves for the
18 bands listed in Table \ref{tab:filter_details}. 
These filters were chosen for similarity with previous work in the LoTSS deep fields \citep[e.g.][]{Smith2021}, {\color{black} providing good coverage of the spectrum from the UV to the FIR with which to test how \magphys\ performs in these idealised conditions}.  
Figure \ref{fig:SED_example} shows the filter coverage for an example SED at z = 1, along with the emergent SED generated by SKIRT.

\begin{figure*}
    \centering
    \includegraphics[width=0.49\textwidth]{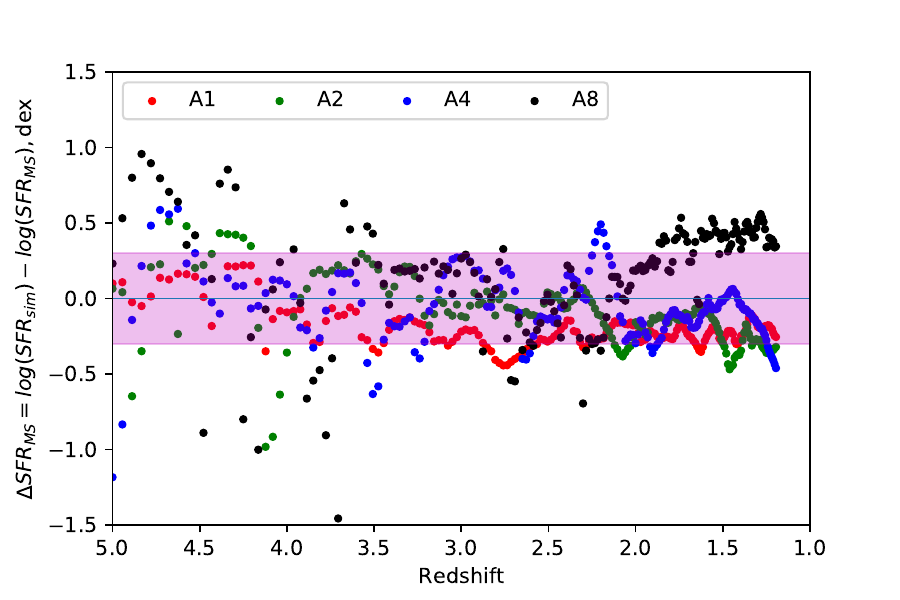}
    \includegraphics[width=0.49\textwidth]{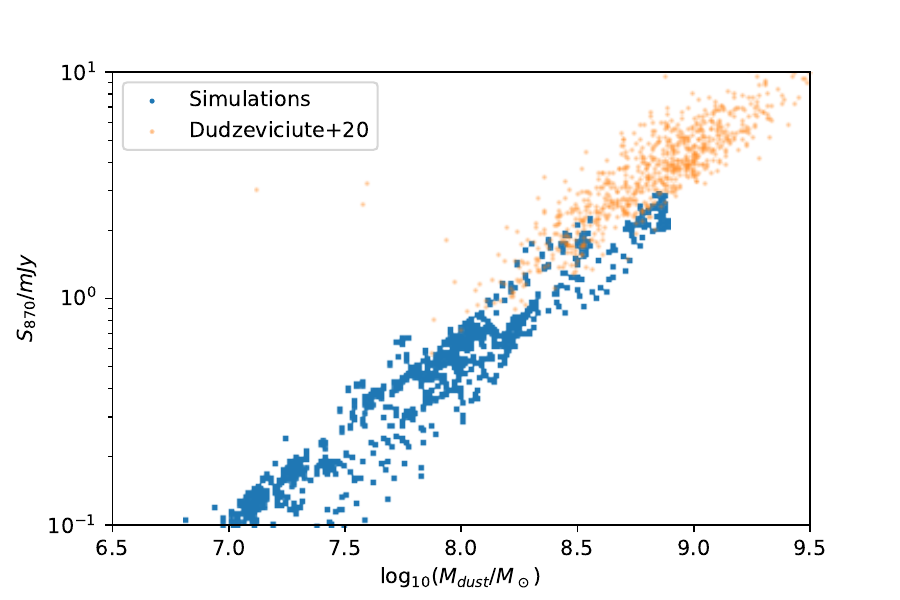}
    \includegraphics[width=0.49\textwidth]{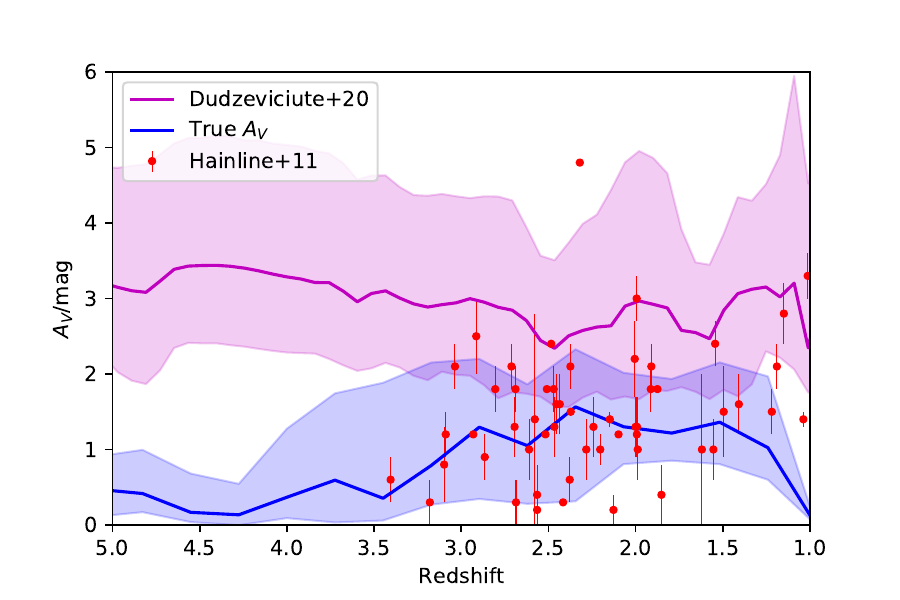}
    \includegraphics[width=0.49\textwidth]{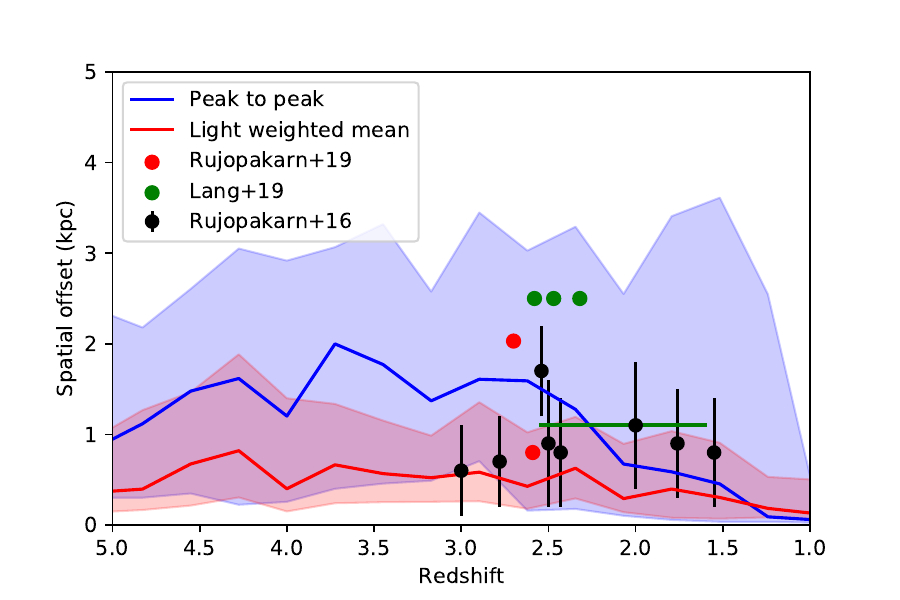}
    \caption{{\color{black}The properties of the simulated galaxies in their observational context. \textbf{Upper left:} the relationship between the simulated galaxies' SFRs and the galaxy main sequence (MS); for each snapshot, the $y$-axis shows the difference between the simulation SFR and the MS, with the magenta band indicating the typical $\pm\,0.3$ dex scatter associated with the MS \citep[e.g.][]{Tacchella2022}. \textbf{Upper right:} the relationship between the sub-mm flux density $S_{870}$ and dust mass; the blue points represent the simulated data, while the orange points show galaxies from D20. \textbf{Lower left:} the variation in $A_V$ as a function of redshift for the simulations (for which the median value at each $z$ is shown by the solid line, within shading  indicating values enclosed by the 16th and 84th percentiles of the distribution), along with a corresponding distribution from D20 (shown in purple). The SMG sample from \citet{Hainline2011} is shown by the red points. \textbf{Lower right:} the mean UV\slash FIR peak to peak (blue) and light-weighted mean (red) spatial offsets in redshift bins: the shading indicates the region enclosed by the 16th and 84th percentiles at each redshift, while the solid line indicates the median value. The red, green and black circles are values for individual sources taken from the literature (as indicated in the legend), while the solid green line marks the reported average spatial offset across 20 SMGs from \citet{Lang2019}.} }
    \label{fig:MSSFR}
\end{figure*}

{\color{black} Figure \ref{fig:MSSFR} examines the relationship between the properties of our simulated galaxies and those of high redshift sub-millimetre galaxy populations in which spatial UV--FIR offsets have been observed. We compared four properties with observations, specifically the SFR relative to the galaxy main sequence (MS; upper left panel), the relationship between sub-mm flux density and $M_{\mathrm{dust}}$ (upper right), the degree of $V$ band extinction (lower left), as well as the magnitude of the UV\slash IR offsets (lower right) in relation to studies in the literature. In the upper left panel we have compared the SFR in each snapshot with the MS parameterisation from \citet{Schreiber2015} modified for our adopted \citet{Chabrier2003} IMF using the method of \citet{Madau2014}, as a function of redshift. 
The magenta band indicates the typical $\pm\,0.3$ dex scatter associated with the MS \citep[e.g.][]{Tacchella2022}. The simulated galaxies lie either on or above the MS in the vast majority of cases, and are therefore consistent with dusty, star forming galaxies. }
{\color{black} The upper right panel of Figure \ref{fig:MSSFR} shows the sub-millimetre flux density, $S_{870}$, as a function of the dust mass for the simulated galaxies and for the SMGs published in D20. While the simulations do not occupy the parameter space of the brightest SMGs, there is significant overlap, and they do lie along the same submm/dust mass relationship (see \citealt{Hayward2011}, \citealt{Cochrane2023}). The lower left panel shows how the $V$-band extinction ($A_V$) for the simulations (the blue solid line indicates the median, with shading indicating the values enclosed by the 16th and 84th percentiles of the distribution at each redshift) compares with the corresponding values for the SMG samples from D20 (in purple) and \citet[indicated by the red points with error bars]{Hainline2011}. Although the D20 sample is on average more obscured than our simulations, similarity to the \citet{Hainline2011} SMGs is evident.  
The lower right panel shows the range of offsets between the UV and FIR emission in redshift bins. The solid lines indicate the mean simulated offset (blue for peak-to-peak, red for light-weighted mean), with shaded regions indicating the area enclosed by the 16th and 84th percentiles at each redshift. The black, red and green symbols indicate ALMA sources from \citet{Rujopakarn2016},  
and \citet{Rujopakarn2019} and \citet{Lang2019}.
Finally, the short green line marks the mean offset from \citet{Lang2019} over 20 SMGs with $1.6<z<2.5$.  
\par 
To summarise, Figure \ref{fig:MSSFR} demonstrates that the simulated sources are predominantly dusty star-forming galaxies. While the D20 SMG sample is more extreme, the degree of extinction and the magnitude of the UV-FIR spatial offsets in the simulations show significant overlap with values published in the literature. The simulations are therefore a useful testing ground for determining the extent of our ability to recover the true properties of galaxies with plausible UV--FIR offsets using \magphys. 

}

\begin{table}
 \centering
 \caption{The filters used to create synthetic observations from the simulated photometry. The first column gives the telescope\slash survey, the second the instrument\slash filter name, and the third the effective wavelength of the filter. }
 \label{tab:filter_details }
\begin{tabular}{lll}
\cline {1-3}
Facility & Filter & $\lambda_\mathrm{eff}(\mu $m)\\
\cline {1-3}
CFHT&Megacam $u$&0.39\\
PanSTARRS&$g$&0.48\\
PanSTARRS&$r$&0.61\\
PanSTARRS&$i$&0.75\\
PanSTARRS&$z$&0.87\\
PanSTARRS&$y$&0.96\\
UKIDSS & $J$ & 1.2\\
UKIDSS & $K$ & 2.2\\
\textit{Spitzer} &IRAC ch1& 3.4 \\
\textit{Spitzer} &IRAC ch2& 4.5\\
\textit{Spitzer} &IRAC ch3& 5.6\\
\textit{Spitzer} &IRAC ch4& 8.0\\
\textit{Spitzer} &MIPS 24\,$\mu$m & 24\\
\textit{Herschel} &PACS green& 100 \\
\textit{Herschel} &PACS red& 160\\
\textit{Herschel} &SPIRE & 250 \\
\textit{Herschel} &SPIRE & 350 \\
\textit{Herschel} &SPIRE & 500 \\
\cline {1-3}
\end{tabular}
\label{tab:filter_details}
\end{table}

\color{black}
\subsection{\magphys}
\label{sec:{MAGPHYS}}
 \magphys\ is an SED modelling code using Bayesian inference to derive best-fit SEDs as well as estimates (best-fit, median likelihood, and probability distribution functions) for a wide range of galaxy properties. A full description can be found in DC08 and \citet{daCunha2015}, but we include a brief overview. \magphys\ uses two libraries of model galaxies: the first, the library of star-formation histories (SFH), consists of 50,000 models each comprising a UV/optical SED and associated galaxy properties; the second, the dust library, comprises 25,000 models each with an IR SED and associated properties.
\par
The SFH library is built using the IMF of \citet{Chabrier2003} and the stellar population synthesis (SPS) model of \citet{Bruzual2003}. Exponentially declining star formation histories are superposed with random bursts, in such a way that a burst of star formation has occurred in half of the SFH library models within the last 2$\,$Gyr.

Common to both libraries is the use of the \citet{Charlot2000} two-component dust model. In this model, stellar populations younger than 10\,Myr are attenuated by a greater amount than older stellar populations, under the assumption that these young stars are still embedded within their `birth clouds'. These stellar populations are subject to a total optical depth $\tau_{\rm BC} + \tau_{\rm ISM}$,
whereas older populations `see' an optical depth of only $\tau_{\rm ISM}$, from the diffuse ISM. {\color{black} \citet{Charlot2000} define the optical depth seen by stellar emission as
\[
\hat{\tau_\lambda} = 
\begin{cases}
\hat{\tau}_\lambda^{BC}+\hat{\tau}_\lambda^{ISM}&\text{for stars < $10\,$Myr},\\
\hat{\tau}_\lambda^{ISM}&\text{for stars $\geq 10\,$Myr}.\\
\end{cases}
\]

\noindent where $\hat{\tau}_\lambda$ is the total optical depth for $\lambda$, $\hat{\tau}^{BC}_\lambda$ is the optical depth of the birth clouds and $\hat{\tau}^{ISM}_\lambda$ is the optical depth of the ISM. These latter two are defined in \magphys\ such that:
\begin{align}
    &\hat{\tau}^{BC}_\lambda = (1-\mu)\hat{\tau}_V(\lambda/5500\text{\r{A}})^{-1.3},\, \mathrm{and} \label{eqn:dust1}\\
    &\hat{\tau}^{ISM}_\lambda = \mu\hat{\tau}_V(\lambda/5500\text{\r{A}})^{-0.7},\label{eqn:dust2}
\end{align}
\noindent where $\hat{\tau}_V$ is the mean $V$ band optical depth and $\mu$ represents the fraction of $\hat{\tau}_V$ arising from the ISM.}

\begin{figure*}
    \centering
    \includegraphics[width=0.49\textwidth]{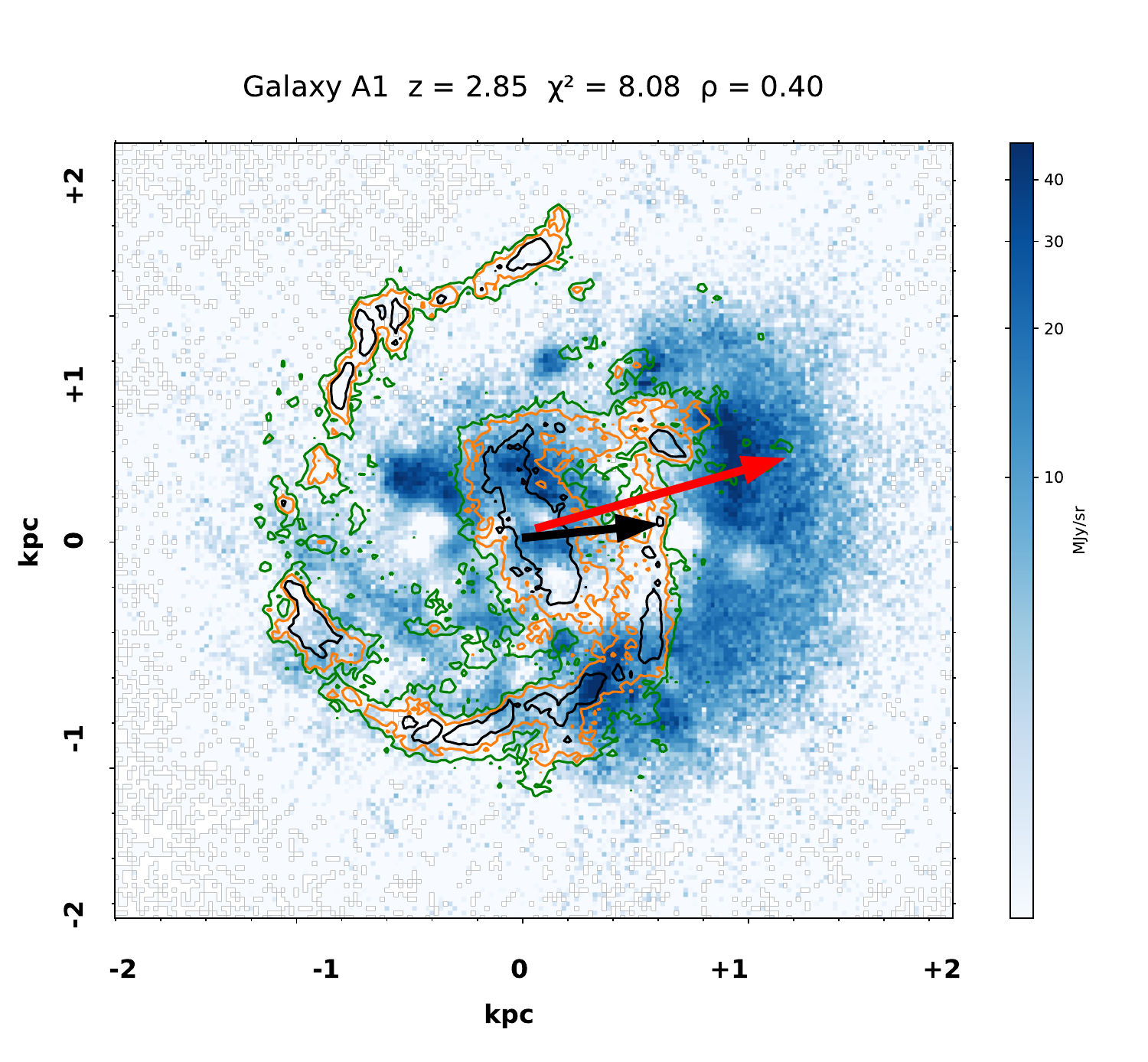}
    \includegraphics[width=0.49\textwidth]{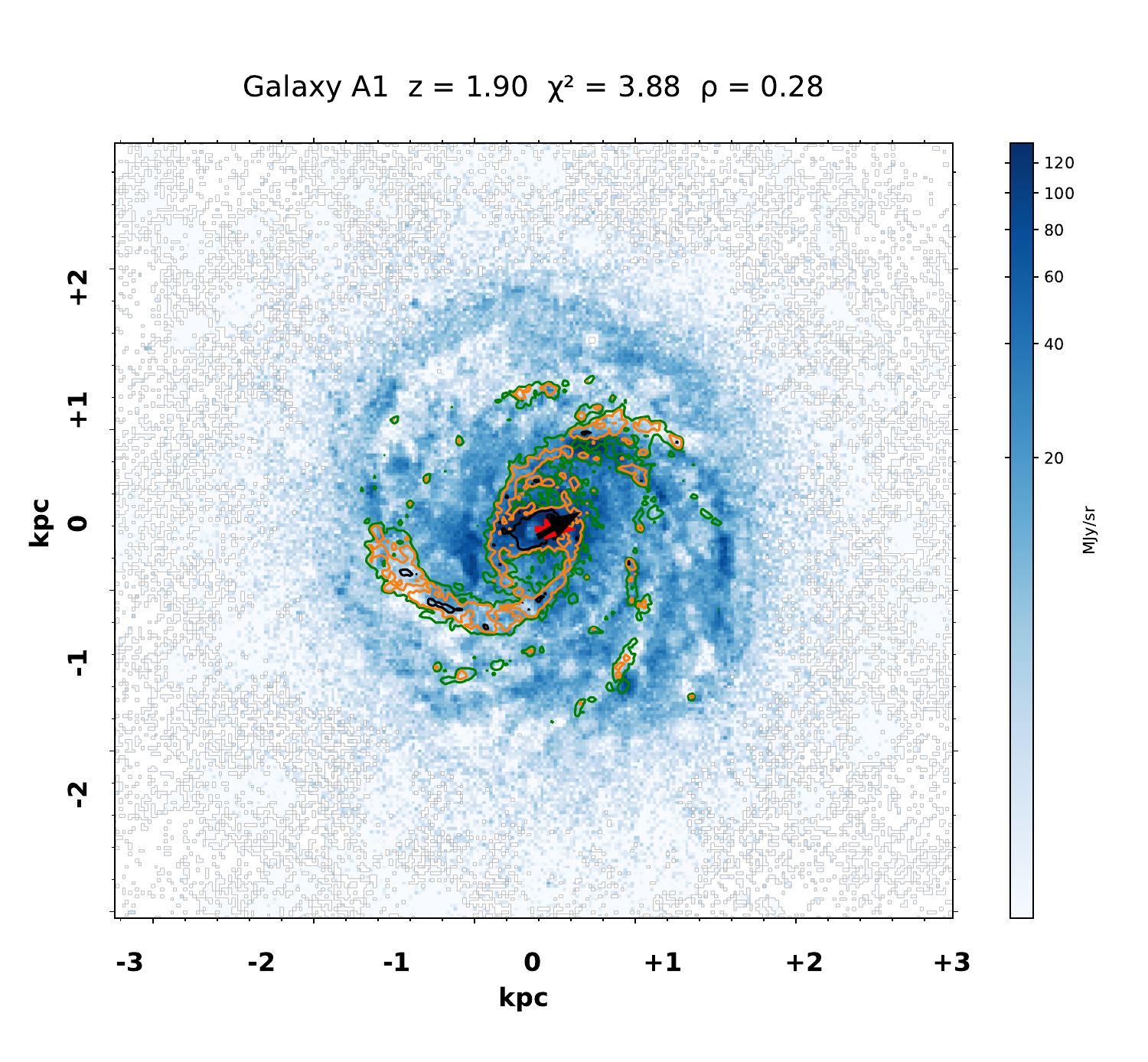}
    \caption{{\color{black}Visualisations of two views of galaxy A1, in the later stages of its evolution, showing} differing degrees of UV--FIR offsets ranging from kpc-scale projected separation (left) to approximately co-spatial (right). In each panel, the image in blue shows the UV emission, the side colourbars showing the flux density of the emission in MJy/sr. The coloured contours show flux density for the FIR emission, ranging from green ($3\times10^4$\,MJy/sr), to
    orange ($5\times10^4$\,MJy/sr) to black ($10^5$\,MJy/sr). In each panel, the base of the red vector is positioned at the peak FIR emission and the head at the peak UV emission, the base of the black vector is positioned at the light-weighted mean FIR emission and the head at the light-weighted mean UV emission. The title of each plot gives the galaxy name along with redshift, best-fit $\chi^2$ and Spearman $\rho$ value. } 
    \label{fig:offset_examples}
\end{figure*}

The dust library is built from three main components: emission from very small grains ($< 0.01\,\mu$m) which can reach high temperatures if they absorb a UV photon; large grains (between $0.01\,-\,0.25\,\mu$m) in thermal equilibrium with the interstellar radiation field; and polycyclic aromatic hydrocarbons (PAHs) which are responsible for emission line features in the mid-infrared. The contribution of each component to the SEDs of the birth clouds and the ISM is chosen to broadly reproduce the range of SEDs found in nearby star-forming galaxies. The total IR SED is then modelled as the sum of the ISM and the birth cloud components.

The SFH and dust libraries are linked together in such a manner that the starlight absorbed by dust at short wavelengths is re-radiated at longer wavelengths, i.e. the energy is balanced. 
During the fit, as well as ensuring that energy conservation (i.e.\, energy balance) is satisfied by construction (i.e.\,the luminosity absorbed by dust equals that emitted by dust), \magphys\ combines those models in the optical library with those in the IR library that have similar contributions from dust in the ISM to the overall dust energy budget (the fraction of luminosity absorbed by the diffuse ISM
component and that emitted by the diffuse ISM component, respectively).
This is parameterised in \magphys\ using the $f_\mu$ parameter; in the high-redshift version used in this work, values for the SFH and dust libraries must have $\Delta f_\mu < 0.2$ for the combination to be acceptable. 
In this way, each galaxy is fitted against a wide variety of `empirical but physically-motivated' (DC08) SFHs and dust content. 
By calculating the best-fit $\chi^2$ for each model combination that satisfies the conditions, a likelihood function is built for each galaxy property by assuming that $L\propto \exp{(\frac{-\chi^2}{2})}$. 
When all combinations of models in the libraries have been processed, a PDF is produced for each property by marginalising the individual likelihoods. \magphys\ outputs a pair of files for each fitted galaxy: one containing the best-fit SED (an example of both the attenuated and unattenuated versions are shown, alongside the model photometry in Figure \ref{fig:SED_example}), while the other contains the best-fit model values and the PDFs.

This study uses the high-redshift version of \magphys\  \citep{daCunha2015}, which differs from the low-redshift version in two important ways: firstly, the prior distributions are modified to include higher dust optical depths, higher SFRs and younger ages; secondly, the effects of absorption in the inter-galactic medium (IGM) at UV wavelengths are taken into account.

Some studies have sought to determine the extent to which AGN can influence the results of SED fitting (e.g. HS15, Best et al., \textit{in preparation}). However, neither the simulations nor the SED fitting code used in this paper include AGN, and so this important aspect will not be discussed further. 

\subsection{Processing the data}\label{sec:Processing_the_Data}
To test how well \magphys\ is able to recover the intrinsic properties of the simulated galaxies, we ran \magphys\ four times on each synthetic SED, using different combinations of photometry and assumed redshift:
\begin{itemize}
    \item Run A - used all 18 filters;
    \item Run B - used all 18 filters, but with all SEDs shifted to a redshift of 2. This run was used as a comparison to detect any bias in the results due to redshift effects. 
    This is discussed in section \ref{sec:Sampling bias};
    \item Run C - used only the UV to near-IR filters ($u$ -- $K$);
    \item Run D - used only the FIR filters (PACS 100\,$\mu$m -- SPIRE 500\,$\mu$m). 
\end{itemize}
\noindent Runs C \&\ D are discussed in section \ref{Fitting only part of the spectrum}. We assumed a signal-to-noise ratio of 5 in every band, following \citet{Smith2018}.
\par
One of the key aims of this work is to determine how \magphys\ performs when analyzing galaxies for which the observed UV and FIR emission are spatially `decoupled.' To do this, we characterise the offset between the UV and FIR emission in three different ways: 

\begin{enumerate}
    \item the peak to peak offset: this is defined as the distance in parsecs between the points of maximum flux in the UV ($0.3\,\mu$m) and FIR ($100\,\mu$m) images;
    \item the light-weighted mean offset: this is defined as the distance in parsecs between the light-weighted centres for the UV ($0.3\, \mu$m) and FIR ($100\,\mu$m) emission. 
    \item the Spearman rank coefficient \citep{myers2003} comparing the degree of correlation between the UV ($0.3\,\mu$m) image and the FIR ($100\,\mu$m) image. A Spearman rank coefficient of $\rho>0.8$ is considered necessary for a strong correlation. Spearman also returns a $p$ value indicating a correlation confidence level, 99 per cent of our results returned $p$ values indicating that the probability of the reported correlation being due to chance was $<0.0001$. The images were filtered to allow only the data points with intensity above the 80th percentile in either the UV or FIR images to be included in the analysis. This was done to avoid the comparatively very large number of low intensity pixels from unduly dominating the result. The 80th percentile was chosen as a reasonable value after comparing the results using different percentile values of the UV and FIR images by eye.
\end{enumerate}

The three proxies were each calculated using the rest-frame UV and FIR maps for each snapshot and view to provide values that would be possible using real observational data with high enough spatial resolution and sensitivity. {\color{black}As an example, Figure \ref{fig:offset_examples} shows two images of the simulated galaxy A1 in the later stages of its evolution, other examples can be seen in \citealt{Cochrane2019}.}  The image on the left shows a significant offset between the UV (shown as the blue image) and FIR (shown as contours) intensity, while in the right image (which has the same colour scheme) the UV and FIR appear almost coincident. In both panels the red vectors show the peak-to-peak offset, while the black vectors show the light-weighted offset. The Spearman $\rho$ value is given in the title of each panel. We also calculated the offsets using the projected maps of the simulated young stars (age $< 10$\,Myr) and dust; however, there was no significant difference in the results and so the observed offsets are used throughout this paper.

In the following sections, where we compare derived values to true (simulated) values these are expressed as residuals in dex between the 50th percentile of the derived value's likelihood function and the true value: 
\begin{equation}\label{eqn:residual}
    \Delta \log (\mathrm{parameter}) = \log_{10}(\mathrm{derived\ value})-\log_{10}(\mathrm{true\ value}).
\end{equation}

\noindent It follows that positive offsets ($\Delta$) represent \magphys\ over-estimates, and negative values  indicate under-estimates. Throughout this work, where \magphys\ results are shown averaged across the seven views of a snapshot, they are the mean of the individual median likelihood estimates. 

\section{Results}
In this section we present results from the four  runs described in Section \ref{sec:Processing_the_Data}. In all runs a successful fit was defined as one where the $\chi^2$ value was equal to or below the 99 per cent confidence limit ($\chi^2_{\mathrm{max}})$, this was taken from standard $\chi^2$ tables. {\color{black} The number of degrees of freedom was calculated as in \citet{Smith2012b}, which perturbed the output best-fit SEDs from \magphys\ with random samples from the standard normal distribution and found that it depended on the number of bands in the manner shown in Appendix B of that work. We are using the same \magphys\ model and have assumed that the relation does not vary with the particular choice of bands or the redshifts of the sources being studied.}

\subsection{The fraction of mock observations with acceptable fits}
\label{sec:Success rate for fitting an SED}

From run A we find that \magphys\ achieved a statistically acceptable fit (i.e. $\chi^2 \le \chi^2_\mathrm{max}$) for 83\, per cent (5,567 out of 6,706) of the snapshots. {\color{black}Note that the value of $\chi^2_{\mathrm{max}}$ varies with redshift because the SKIRT SEDs do not include wavelengths $<0.08\,\mu$m, meaning that we are unable to generate synthetic photometry for the bluest filters at $z \gtrsim 3.9$.} 

\begin{figure}
\centering

\includegraphics[width=0.46\textwidth]{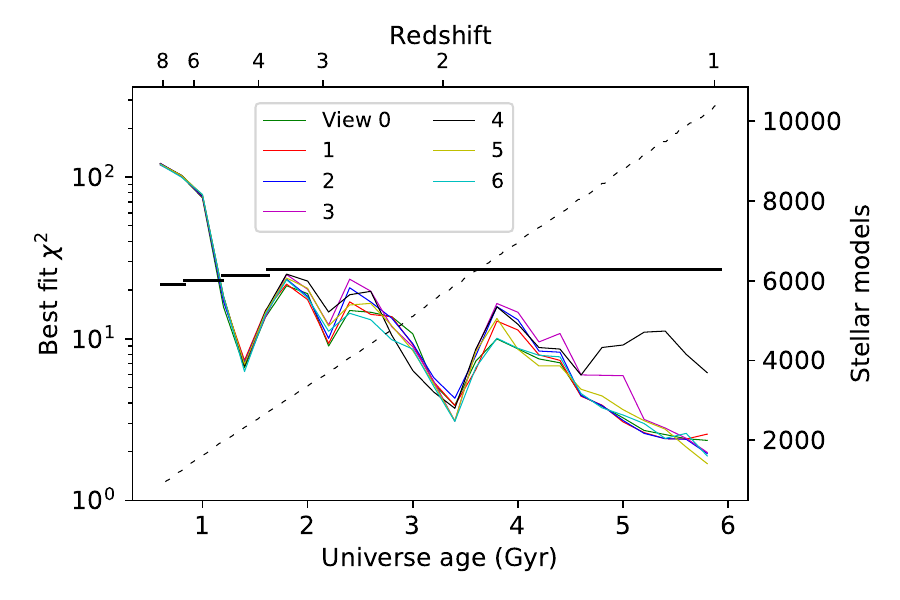}
\caption{\magphys\ produces statistically acceptable fits for virtually all snapshots at $z < 5$, irrespective of viewing angle. The best-fit $\chi^2$ as a function of Universe age is shown for galaxy A1, colour-coded by view number. The $\chi^2$ values have been averaged over bin widths of $\Delta z = 0.2$ (relative to the top horizontal axis) for clarity. The horizontal line indicates the $\chi^2$ threshold below which a fit is deemed acceptable using the \citet{Smith2012b} criterion{\color{black}, this value varies with redshift (see Table \ref{tab:chi2values})}. The dashed line indicates the number of stellar models (relative to the right-hand $y$-axis) available to \magphys\ at a given redshift with which to compare the input SED.
Although not shown here, qualitatively similar results are obtained for the other simulations (A2, A4 \&\ A8).}
\label{fig:chi2_success_by_view}
\end{figure}

\begin{figure}
\centering
\includegraphics[width=0.47\textwidth]{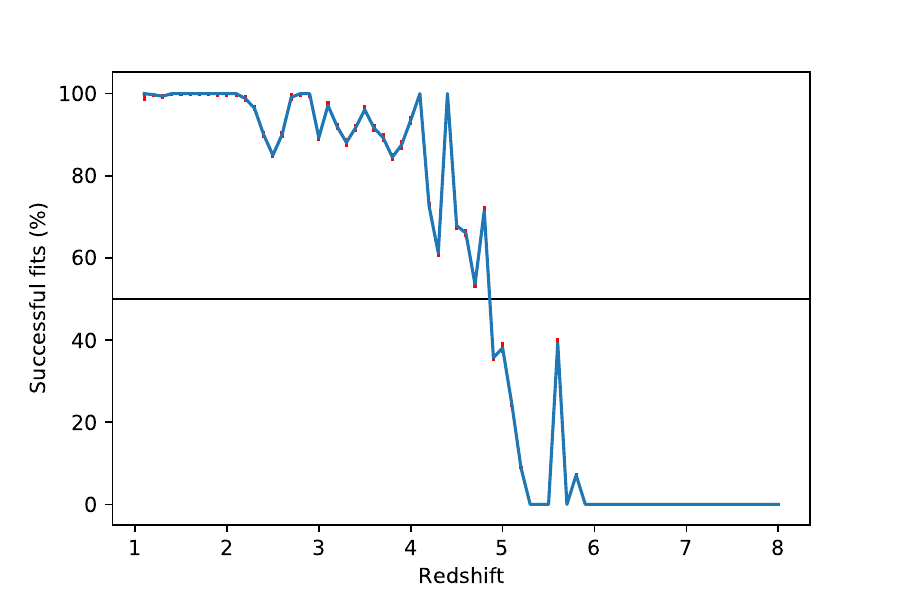}
\caption{\magphys\ success rate in fitting SEDs. The percentage of successful fits averaged across all views and snapshots of all galaxies as a function of redshift, note that standard Poisson errors are too small to be visible. The horizontal line marks a success rate of 50 per cent. The fraction of fits that are statistically acceptable decreases with increasing redshift due to the constraint that the SFH must be shorter than the age of the Universe at that redshift, meaning that the size of the template library decreases with increasing redshift.}
\label{fig:chi2_success_rate}
\end{figure}

\begin{table}
\caption{{\color{black}The number of filters available and the value of $\chi^2_{\mathrm{max}}$ for different redshift ranges. }}
\centering
\begin{tabular}{cccc}
\\
\cline {1-3}
z&filters&$\chi^2_{\mathrm{max}}$\\
\cline{1-3}
$8.4\leq z<6.6$&15&21.67\\
$6.6\leq z<5.0$&16&23.01\\
$5.0\leq z<3.9$&17&24.75\\
$3.9\leq z<1.0$&18&26.72\\
\cline{1-3}
\end{tabular}
\label{tab:chi2values}
\end{table}
The derived $\chi^2$ values are broadly independent of viewing angle for all galaxies; as an example, Figure \ref{fig:chi2_success_by_view} shows the $\chi^2$ results for all snapshots and views for the galaxy A1. Figure  \ref{fig:chi2_success_rate} shows how the fit success rate, averaged across all snapshots and views for all four galaxies, changes with redshift. We see from this that \magphys\ can routinely produce acceptable fits to the synthetic photometry up to $z=4$, but that the success rate drops to 50 per cent at $z \approx 4.85$ and to zero after $z \approx 5.9$. Different factors may be contributing to this effect. Firstly, the number of SFHs from the \magphys\ libraries that are compared with observations is a strong function of redshift. \magphys\ does not consider SFHs longer than the age of the Universe at a given redshift (the number of SFHs shorter than the Hubble time at each redshift is shown as the dashed line, relative to the right-hand axis in Figure \ref{fig:chi2_success_by_view}) and 
at z $\approx$ 5 the number of such SFHs in the library is only 20 per cent of those available at z $\approx$ 1. It is therefore clear that the prior is significantly more densely sampled at lower redshifts, leading to more acceptable fits in cases such as this, where the SFH itself is constrained only weakly by the photometry \citep[e.g.][]{Smith2018}.
Secondly, at these very early times in the simulations ($z>5$), the model galaxies are low mass ($< 10^9$\,M$_\odot$) and bursts of star formation have a disproportionate influence on a galaxy's bolometric luminosity. This highly stochastic star formation is not well-modelled by the star formation histories included in the \magphys\ libraries. {\color{black} It is possible that including additional bands of model photometry may provide better results, e.g. by an additional sub-millimetre datapoint providing an `anchor' point to the Rayleigh-Jeans tail of the dust SED and in doing so enabling tighter constraints on the overall energy balance (though we note that the 500\,$\mu$m band does sample this side of the dust SED out to $z \approx 4$). However, in this work we have chosen to focus on an example set of photometric data appropriate for studying dusty star-forming galaxies in general, and with an enforced SNR = 5 in every band we are not subject to some of the sensitivity (or resolution) limitations associated with using real \textit{Herschel} data to study galaxies at the highest redshifts. We therefore defer testing our results with different photometric coverage for a future work.} 
Throughout the remainder of this study, we follow the same approach used in previous \magphys\ works both observational and numerical \citep[e.g. HS15;][]{Smith2012b,Smith2018,Smith2021}, and consider only those views for which an acceptable fit was obtained.

To investigate the influence of redshift on the \magphys\ fit rate further, we used Run B, in which the photometry is modified such that all SEDs were placed at $z=2$.  In this run, the size of the libraries and therefore the sampling of the priors used for SED fitting is the same for all snapshots. We find that the fit success rate increases to 93\,per cent for the forced $z=2$ runs, from 83\,per cent for run A. Although it is tempting, we cannot attribute this change solely to the weakening of the SFH prior, since it is also possible that sampling different rest-frame wavelengths could impact the fit success rate \citep[e.g. because of individual spectral features being redshifted into a particular observed bandpass;][]{Smith2012b}. These effects are discussed further in section \ref{sec:Sampling bias}.

\subsection{Overall \magphys\ performance}\label{sec:Recovery of galaxy properties}

In studying the fidelity of the \magphys\ parameter estimates, we have chosen to focus on five properties likely to be of the widest interest, namely SFR and sSFR (both averaged over the last 100$\,$Myr), $M_{{\mathrm{star}}}$,  $M_{{\mathrm{dust}}}$ and $L_{{\mathrm{dust}}}$. The true values for $M_{{\mathrm{star}}}$, SFR (averaged over the last 100$\,$Myr), and $M_{{\mathrm{dust}}}$ were available from the simulation. The true values for $L_{{\mathrm{dust}}}$ were calculated by integrating under the {\sc{SKIRT}}-produced rest frame SED from $8\,\mu$m$<\ \lambda < 1000\,\mu$m, following \citet{Kennicutt1998a}. 

\subsubsection{The fidelity of \magphys\ results over time}

\begin{figure*}
\centering
\includegraphics[width=0.9\textwidth]{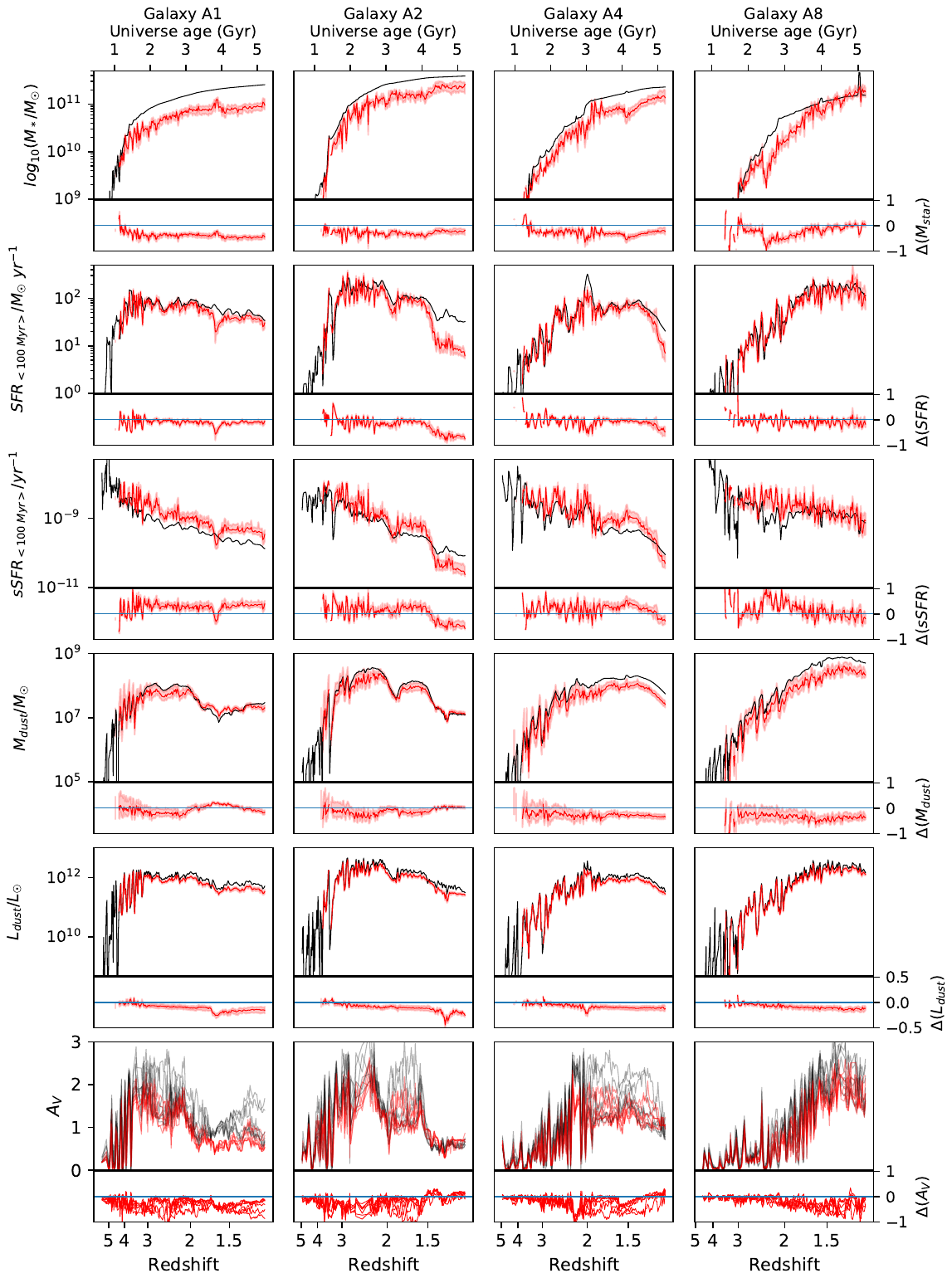}
\vspace{-1.5 cm}
\caption{The overall \magphys\ parameter estimation (red) compared with the true values from the simulation (black); \magphys\ captures the overall true properties as a function of redshift. The columns refer to galaxies A1, A2, A4 and A8, respectively. In each row, the upper plot presents the evolution against universe age (upper $x$-axis) and redshift (lower $x$-axis), and the lower plot shows the residuals on the same $x$-axes (note that the range for $\Delta L_{\mathrm{dust}}$ is smaller than that for other properties). The top row presents the evolution of stellar mass, {\color{black}while the four subsequent rows present the corresponding evolution of SFR, sSFR, $M_{{\mathrm{dust}}}$ and $L_{{\mathrm{dust}}}$ respectively. In each main panel, the black line indicates the true values, the red line plots the mean across all views of the median recovered value, and the shaded area indicates the region enclosed by the typical error bar on each parameter (i.e. the mean difference between the 16th\slash 84th percentile and the median, for the upper and lower bounds, respectively). In the final row, the black and red lines in the upper plot show the true and recovered values of $A_V$ for the different views, while the lower plot shows the residuals for each view.}  }
\label{fig:Evolution_of_parameter_recovery}
\end{figure*}

Figure \ref{fig:Evolution_of_parameter_recovery} shows the evolution in the true and derived physical properties of our simulated galaxies as a function of redshift (with a second horizontal axis at the top of each column showing the age of the Universe at each redshift in our adopted cosmology). The different physical properties are shown along successive rows, while the different simulated galaxies are shown in successive columns, as indicated in the text at the top of each column. In each panel, the black line indicates the true values for each property, taken from the simulations, while the red line indicates the mean of the median-likelihood \magphys\ estimates, where the averaging has been conducted over the seven different viewing angles. Similarly, the shaded red region in each panel indicates the area enclosed by the mean of the 16th and 84th percentiles of each parameter's \magphys\ PDF (once more averaged over the seven views), to give the reader a feel for the typical error bar. Each lower panel shows the residual, e.g. $\Delta \log$ (SFR), as defined in Equation \ref{eqn:residual}.
\par 
In general, \magphys-derived values show a significant degree of consistency, both in the temporal sense and by comparison to the true values. The temporal sense is a valuable test in its own right as, although \magphys\ fits each snapshot independently, the true values shown in Figure \ref{fig:Evolution_of_parameter_recovery} mostly vary smoothly with time. That this is reflected in the \magphys\ estimates once the error bars are taken in to account, offers broad encouragement for the use of \magphys\ with observational data. 
\par 
Below, we discuss the degree of fidelity in the \magphys\ parameter estimates overall by comparing with the true (simulated) values. It is clear based on even a cursory inspection of the trends visible in Figure \ref{fig:Evolution_of_parameter_recovery} that the \magphys\ estimates have broadly captured the behaviour visible in the true parameter values, such as increasing stellar mass and generally decreasing sSFR. Similar encouragement was found in the earlier work of HS15, though we now extend this to higher-redshift, dustier galaxies for the first time with a sample of very high-resolution simulations. The mean residuals, $\Delta \log($parameter$)$, averaged over the full evolution of each simulated galaxy, are shown in Table \ref{tab:average_residuals}.

\begin{table*}
\caption{Mean residuals -- $\Delta\log({\mathrm{parameter})}$, as defined in Equation \ref{eqn:residual} -- for each property for each galaxy and the average across all galaxies; a negative value indicates an underestimate. The quoted uncertainties indicate the typical uncertainty that \magphys\ derives on that galaxy parameter (equal to half the difference between the 16th and 84th percentiles of the derived PDF).}
\centering
\begin{tabular}{ccccccc}
\\
\cline {1-7}
Galaxy & 
$\Delta \log(M_{{\mathrm{star}}})$&
$\Delta \log(\mathrm{SFR})$&
$\Delta\log($sSFR$)$&
$\Delta \log(M_{{\mathrm{dust}}})$&
$\Delta \log(L_{{\mathrm{dust}}})$&
{\color{black}$\Delta A_V$}\\
\cline{1-7}
A1 & -0.37 $\pm$ 0.08 & -0.10 $\pm$ 0.06 & 0.26 $\pm$ 0.12 & -0.05 $\pm$ 0.14 & -0.10 $\pm$ 0.04& -0.30 $\pm$ 0.07\\
A2 & -0.28 $\pm$ 0.08 & -0.21 $\pm$ 0.07 & 0.06 $\pm$ 0.13 & -0.11 $\pm$ 0.15 & -0.10 $\pm$ 0.04& -0.20 $\pm$ 0.07\\
A4 & -0.27 $\pm$ 0.08 & -0.08 $\pm$ 0.06 & 0.18 $\pm$ 0.13 & -0.26 $\pm$ 0.19 & -0.07 $\pm$ 0.04& -0.20 $\pm$ 0.07\\
A8 & -0.24 $\pm$ 0.09 & -0.05 $\pm$ 0.06 & 0.19 $\pm$ 0.14 & -0.35 $\pm$ 0.21 & -0.08 $\pm$ 0.03& -0.19 $\pm$ 0.07\\
\cline{1-7}
Mean&-0.29 $\pm $0.09 & -0.11 $\pm$ 0.06 & 0.18 $\pm$ 0.13 & -0.19 $\pm$ 0.17 & -0.09 $\pm$ 0.04& -0.22 $\pm$ 0.07\\
\cline{1-7}
\end{tabular}
\label{tab:average_residuals}
\end{table*}

Averaging the results across all views of all snapshots of all galaxies, we find that the stellar mass is typically underestimated by \magphys, recovered with a mean residual of $\Delta \log(M_\mathrm{star}) =-0.29\pm0.09$. This $3.22 \sigma$ result covers a wide range of simulated scenarios, ranging from the early stages of formation, through periods of starburst, tidal disruptions and merger events. By way of comparison, in HS15 the stellar mass was recovered to within 0.2\,dex (which was also the typical uncertainty in that work) for the vast majority of snapshots, across both the isolated disk and major merger simulations. The principal exception to this excellent recovery being a 0.4\,dex underestimate of the stellar mass during the peak period of AGN activity (which we do not simulate here). 
D20 also reported a larger systematic underestimation of stellar mass, with a deviation of $-0.46\pm0.10$\,dex;  our  results therefore fall between those of these two previous studies. {\color{black} We suggest two factors which may be contributing to this systematic underestimation of the stellar mass. Firstly, a sub-optimal choice of SFH (such as we know we have made in this work, since we can see that the simulated galaxies do not have parametric SFHs in Figure \ref{fig:Evolution_of_parameter_recovery}) has been shown to produce biased results \citep{Carnall2019} and in particular an underestimate for stellar mass when applied to star forming galaxies \citep{Mitchell2013,Michalowski2014}. Secondly,  \citet{Mitchell2013} and \citet{Malek2018} have shown that the choice of attenuation law has an impact on the estimation on stellar mass (and it is also clear that the two-component geometry assumed by \magphys\ is not consistent with the ground truth in the simulations where the radiative transfer calculates the attenuation due to ISM dust \textit{in situ}).}

In the second row of Figure \ref{fig:Evolution_of_parameter_recovery}, we show that the \magphys\ SFRs for our simulated galaxies are typically accurate to within $\Delta \log($SFR$) =-0.11\pm0.06$ of the true values ($1.83\sigma$). Of the five properties highlighted in this study, Figure \ref{fig:Evolution_of_parameter_recovery} shows SFR to be the one for which \magphys\ produces perhaps the most accurate reflection of the true values once the uncertainties are considered. However, there are some points of disagreement that are worth mentioning. The first example of this is for galaxy A1 at $z\approx1.7$: this deviation of $\approx-0.59\pm0.16$ dex ($3.7\sigma$) coincides with a local minimum of $M_{\mathrm{dust}}$, perhaps resulting from a strong outflow, and is associated with a brief reduction in the SFR that is not apparent when averaging over 100 Myr. The second example is for galaxy A2 around $1.0 \le z \le 1.5$ at the point where the galaxy has the highest stellar mass ($M_\mathrm{star} > 10^{11}\,$M$_\odot$), and is the most quiescent that we have simulated (sSFR $\approx 10^{-10}$\,yr$^{-1}$). For comparison, HS15 found that SFR was typically recovered to around 0.2-0.3\,dex accuracy\footnote{We note that HS15 compared \magphys\ 100\,Myr-averaged SFRs with instantaneous SFRs rather than values averaged over 100\,Myr, as we do here. Due to the bursty SFHs of the simulated galaxies, these values can differ significantly \citep{Sparre2017,FV2021}. This topic is further discussed below in connection with $A_V$ recovery.}. D20 reported that SFR was typically underestimated by approximately 20\,per cent -- very similar to our value of $\Delta \log($SFR$) =-0.11\pm0.06$\,dex -- attributing this to differences in their     adopted SFHs, dust model and geometry.

The observed effects in sSFR mirror those in stellar mass and SFR as expected. Averaging over all snapshots and views, we obtain a mean offset of $\Delta \log($sSFR$) = 0.18\pm0.13$, a $1.38\sigma$ result which is consistent with the findings of HS15.

Figure \ref{fig:Evolution_of_parameter_recovery} highlights the excellent recovery of the true dust mass; averaging over all snapshots reveals a mean residual of $\Delta \log (M_\mathrm{dust}) = -0.19\pm0.17$ ($1.12\sigma$), suggesting that the results are typically consistent with the true values once the uncertainties are taken into account, consistent with the findings of D20. 

Overall $L_{{\mathrm{dust}}}$ is well recovered with a mean residual of $\Delta \log (L_\mathrm{dust}) = 0.09\pm0.04$; this $2.25\sigma$ result is again in line with the results of HS15. However, {\color{black} the fifth row} of Figure \ref{fig:Evolution_of_parameter_recovery} may suggest a weak trend for a larger $|\Delta \log(L_\mathrm{dust})|$ in the sense that the \magphys\ estimates increasingly underestimate the true values as the simulations progress and the galaxies develop lower sSFR (though note that the scale of the residual panel for $L_\mathrm{dust}$ is half as large as for the other parameters, which exaggerates the size of the effect). It is possible that the assumptions inherent in the two-component dust model used by \magphys, originally optimised to reproduce the observations of local star-forming galaxies (DC08), are no longer appropriate for the high-mass ($M_\mathrm{star} \approx 10^{11}$), highly star-forming  (SFR $> 20$\,M$_\odot\,$yr$^{-1}$) galaxies that are simulated here. 
\par
{\color{black}Finally, while it is not always the case, $A_V$ is in general underestimated, with a mean residual of $\Delta A_V = -0.22 \pm 0.07$ ($3.14\sigma$), similar to the overall fidelity of the stellar mass recovery. This underestimation of the degree of extinction at $V$ band may be linked to the typical underestimation of the overall dust luminosity, though it is interesting to note this does not prevent excellent recovery of the star formation rate for the majority of snapshots. }

\subsubsection{Searching for systematic trends in the \magphys\ fit results}

We used our simulations to determine the consistency of the \magphys-derived galaxy properties across the range of values presented by the simulations. To do this, we binned the residuals defined using equation \ref{eqn:residual} across the full range of each property (stellar mass, SFR, sSFR, dust mass and dust luminosity) from the simulations and plotted the median bin residual. To gauge the significance of our results, we also averaged across all occupants of each bin to calculate the typical uncertainty associated with each \magphys\ fit (although this is by no means constant in our results), and the scatter within each bin. The median residual, typical error bar, and the 16th and 84th percentile values for the scatter were plotted. Systematic trends might be expected to appear as deviations from horizontal lines in these figures; however, our results show that in all cases, the \magphys\ results are remarkably consistent across the full range of values once the two sources of scatter are taken into account, and no further systematic trends can be identified. The plots are shown in Appendix \ref{first appendix}. 

\subsubsection{The importance of panchromatic data in energy balance fitting}\label{Fitting only part of the spectrum}

We now discuss runs C and D, originally mentioned in section \ref{sec:Processing_the_Data}. 
Run C used only the UV-NIR photometry from $u$ to $K$ band ($0.4\,\mu m < \lambda_\mathrm{eff} < 2.2\,\mu m$), while run D retained only the FIR data from the PACS and SPIRE instruments ($100\,\mu m < \lambda_\mathrm{eff}  < 500\,\mu m$). While it is not possible to `switch off' the energy balance criterion in \magphys, runs C and D enable us to make a direct comparison of the results of `traditional' SED fitting (i.e. attempting to recover the stellar mass or dust content of a galaxy from the optical\slash NIR data alone) with both the true values and the full panchromatic run. In both the starlight-only and FIR-only runs, \magphys\ must rely on the physically-motivated model and the energy balance assumption to estimate the properties usually associated with the missing observations (e.g. estimating the dust mass purely on the basis of the observed starlight, or the stellar mass using only FIR data). 

\begin{figure}
    \centering
    \includegraphics[width=0.48\textwidth]{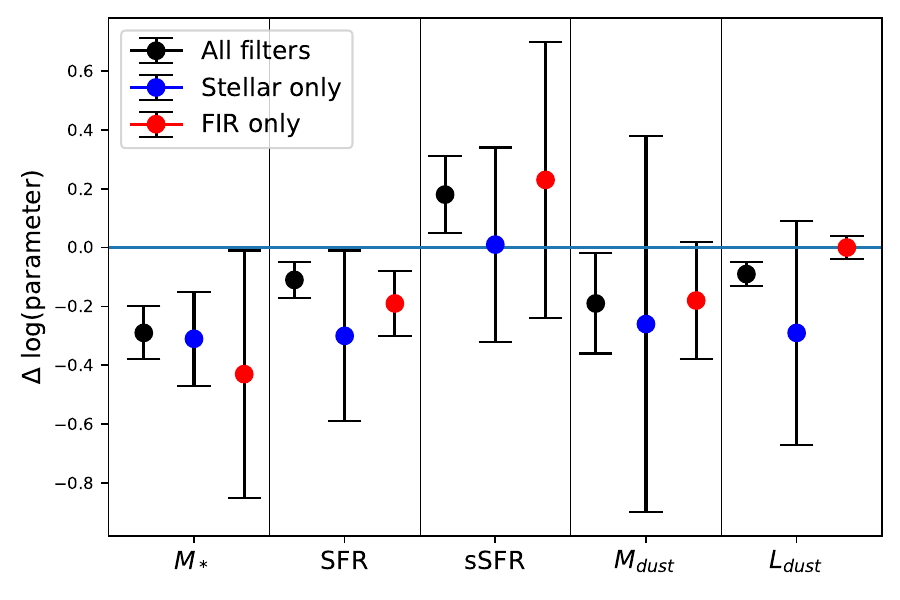}
    \caption{Using \magphys\ to model panchromatic data gives better overall constraints on galaxy properties than sampling only a subset of the available wavelengths. $\Delta \log($parameter$)$ for each parameter of interest, averaged across all galaxies for three different \magphys\ runs: (i) including all available photometry, (ii) stellar only - including only those bands that sample the starlight ($0.4\mu m < \lambda_\mathrm{eff}\,<\,2.2\mu$m), and (iii) FIR only - including only the FIR data ($100\mu$m$ < \lambda_\mathrm{eff}\,<\,500\mu$m), with each set of results colour-coded as in the legend. The error bars on each data point represent the mean uncertainty for each \magphys\ estimate, based on using the 16th and 84th percentiles of the estimated PDFs.}
    \label{fig:residuals_against_runs}
\end{figure}

Figure \ref{fig:residuals_against_runs} shows the results of these runs comparing the mean $\log \Delta$ and typical uncertainty for the five properties for each of the three runs A, C \&\ D: full filter set, stellar-only and FIR-only.

The left panel of Figure \ref{fig:residuals_against_runs} shows the view and snapshot-averaged  $\Delta \log (M_{\mathrm{star}})$ for the three runs. It is immediately clear that although the average $\Delta \log (M_{{\mathrm{star}}})$ is very similar for the stellar-only (0.31\,dex) and all-filter (0.29\,dex) runs, including the full set of data does reduce the typical uncertainty (shown by the error bars) from $\pm 0.20$\,dex to $\pm 0.09$\,dex. Unsurprisingly, attempting to estimate the stellar mass using only the FIR data leads not only to a large $\Delta \log (M_{{\mathrm{star}}})$ but also a significantly larger typical uncertainty ($\approx 0.42$\,dex).

In the second panel, we show the corresponding results for $\Delta \log($SFR$)$. The power of panchromatic fitting is again clear, since the largest $\Delta \log ($SFR$)$ and typical uncertainty occur for the stellar-only fits, which can be influenced by the dominance of the lowest-attenuation sightlines (meaning that the amount of obscured star formation can be underestimated) as well as subject to the well-known age-dust degeneracy \citep[e.g.][]{Cimatti1997}. Our results show that FIR-only SFR estimates are more reliable than those using the $u$ to $K$-band photometry alone, since the FIR-only mean $\Delta \log ($SFR$) \approx 0.19 \pm 0.11$ is significantly closer to the true values than the corresponding stellar-only fits which have $\Delta \log ($SFR$) \approx 0.30 \pm 0.29$. 

The situation is even more pronounced for the recovery of the sSFR, with $\Delta \log ($sSFR$)$ for the three runs shown in the central panel of Figure \ref{fig:residuals_against_runs}. Although the mean $\Delta \log ($sSFR$)$ for the stellar-only run is closest to the true values, the typical uncertainties on the panchromatic run are more than a factor two smaller than the stellar-only estimates. The larger error bar represents a wide range of possible activity levels, making it impossible to unravel the age\slash dust degeneracy; by adding FIR data, the sSFR is better constrained. This, in turn, enables a constrained determination of the SFR and hence the cause of any observed reddening.

For M$_{{\mathrm{dust}}}$, Figure \ref{fig:residuals_against_runs} shows that the addition of stellar data makes very little difference to the mean  $\Delta \log ($M$_\mathrm{dust})$ with FIR-only giving results within 0.18 dex and the full filter set 0.19 dex; this is comparable to the typical uncertainties (0.20\,dex as opposed to 0.17\,dex). Using only the stellar data, the mean $\Delta \log ($M$_\mathrm{dust})$ is 0.26\,dex but the typical uncertainty is significantly increased to 0.64\,dex, reflecting the difficulty associated with estimating the dust content of distant galaxies using data probing the starlight alone. 

Finally, the right-hand panel of Figure \ref{fig:residuals_against_runs} shows the recovery of $L_{{\mathrm{dust}}}$ across the three runs. Interestingly, although the typical uncertainties are similar for the FIR-only and panchromatic runs, the inclusion of the UV/NIR data along with the energy balance criterion perhaps increases the mean $\Delta \log ($L$_\mathrm{dust})$, although the significance of this difference is low.

\subsection{Measuring the effect of UV/FIR `decoupling' on the fidelity of \magphys\ results}

\begin{figure*}
\vspace{-0.5 cm}
\begin{subfigure}{0.43\textwidth}
\includegraphics[width=\textwidth]{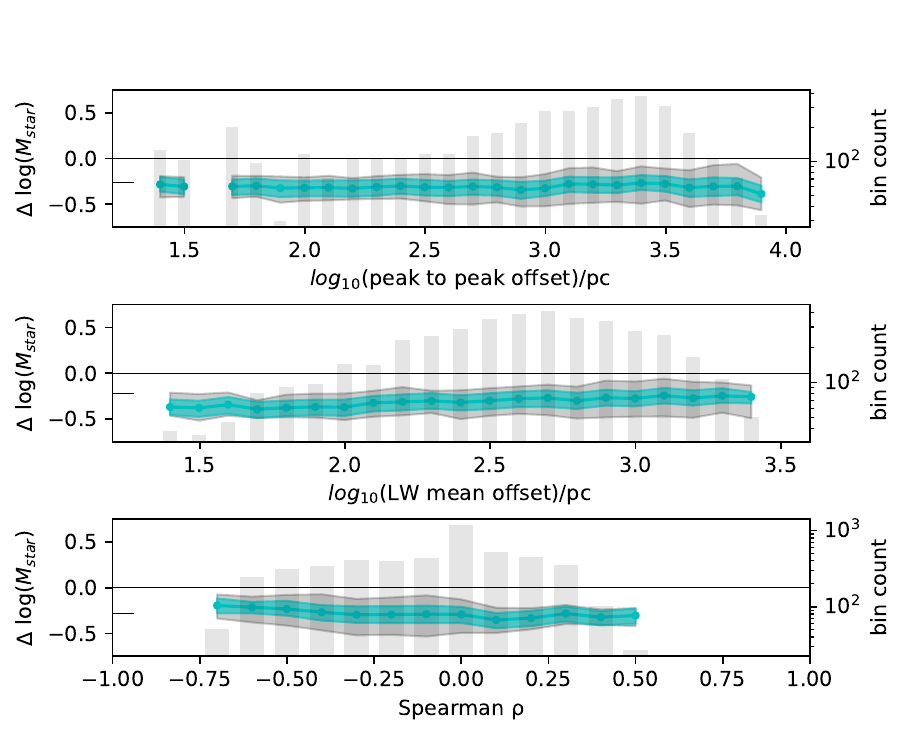}
\caption{$M_{\mathrm{star}}$}
\end{subfigure}
\begin{subfigure}{0.43\textwidth}
\includegraphics[width=\textwidth]{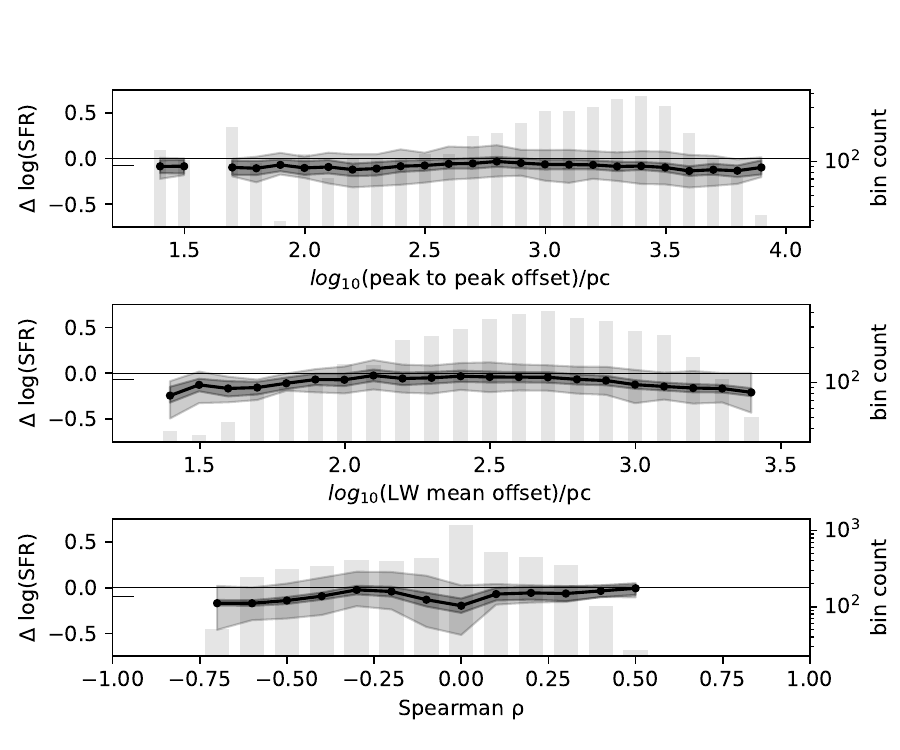}
\caption{SFR}
\end{subfigure}

\begin{subfigure}{0.43\textwidth}
\includegraphics[width=\textwidth]{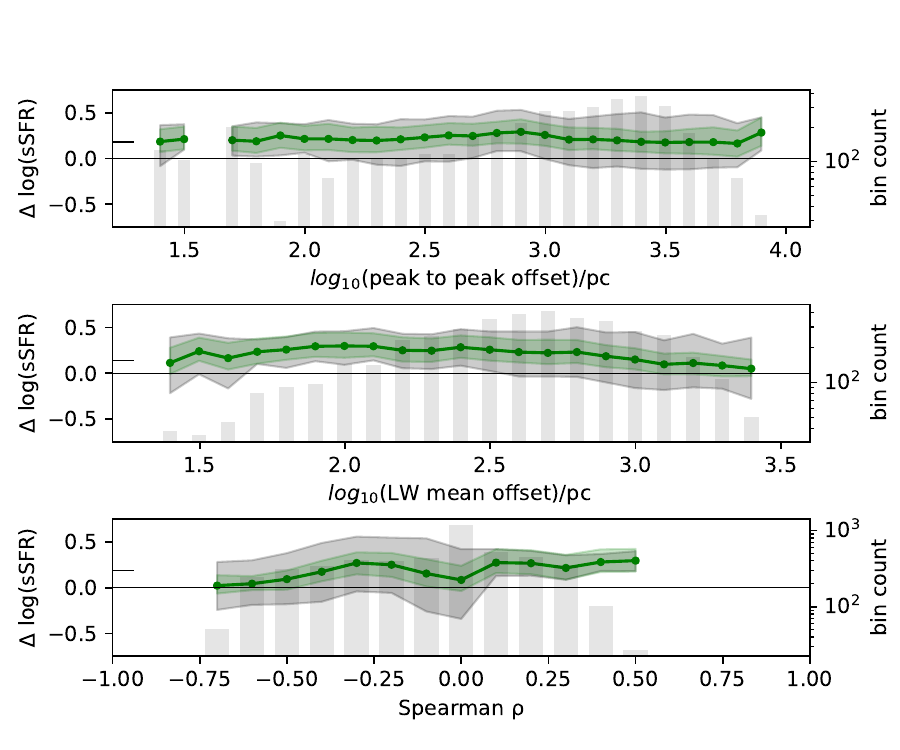}
\caption{sSFR}
\end{subfigure}
\begin{subfigure}{0.43\textwidth}
\includegraphics[width=\textwidth]{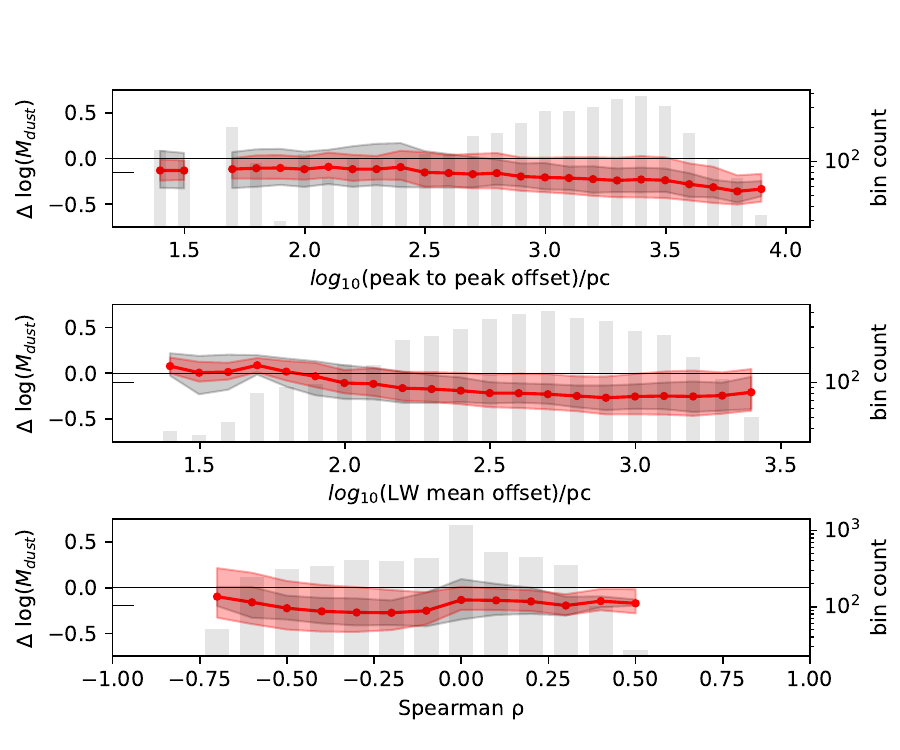}
\caption{$M_{\mathrm{dust}}$}
\end{subfigure}
\begin{subfigure}{0.43\textwidth}
\includegraphics[width=\textwidth]{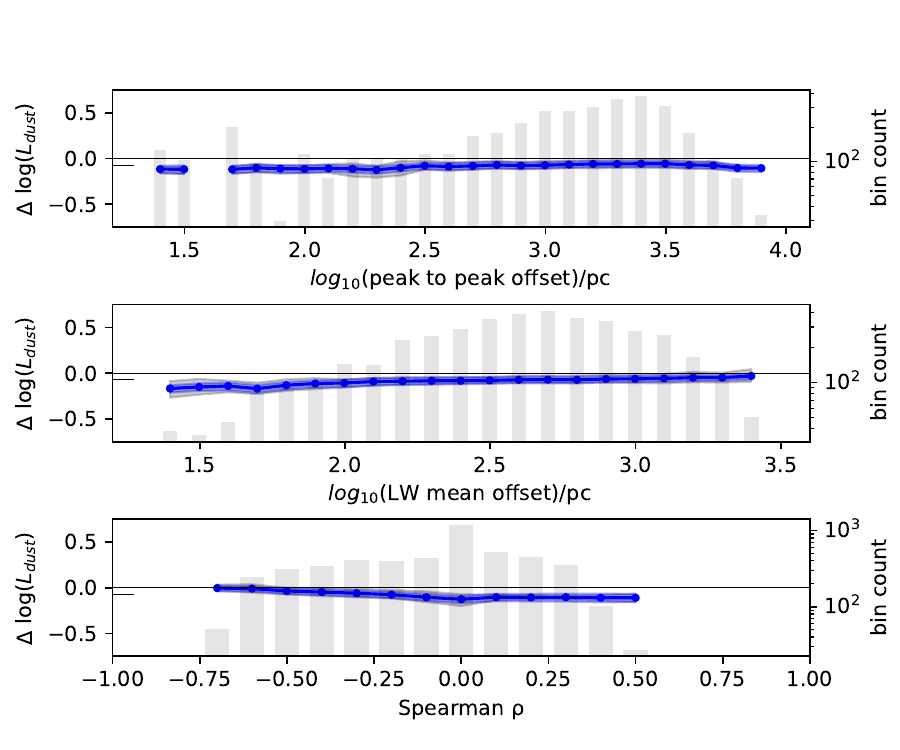}
\caption{$L_{\mathrm{dust}}$}
\end{subfigure}
\vspace{-0.2 cm}
\caption{The fidelity of \magphys\ is largely independent of the extent of any UV\slash FIR offset, as measured by the three proxies, once the uncertainties are considered. $\Delta \log($parameter$)$ as a function of three proxies for the difference between the UV and FIR images - panel (a) presents the data for $M_{\mathrm{star}}$, (b) for SFR, (c) for sSFR, (d) for $M_{\mathrm{dust}}$ and (e) for $L_{\mathrm{dust}}$. For each property, the data points represent the mean over all views and snapshots in that bin. The shaded area of the same colour indicates area enclosed by the mean 16th and 84th percentile values within the bin. The grey shaded area shows area enclosed by the 16th and 84th percentile values for the scatter within each bin. The top plot in each panel shows the logarithmic difference $\Delta$, as a function of the peak-to-peak distance between the UV and FIR images; the second and third panels show the corresponding $\log \Delta$ as a function of the light-weighted mean UV-FIR offset and the Spearman rank correlation coefficient $\rho$ between the 20 per cent brightest pixels in either the UV or FIR images. The short coloured lines adjacent to the left-hand y-axis represent the overall mean value. The grey histograms in each panel (a) to (e) show the bin occupancy relative to the right-hand axis. }
\label{fig:Residuals_against_dust}
\end{figure*}

As discussed above, the primary goal of this work is to examine the fidelity of the \magphys\ results as a function of the degree of correlation or apparent offset between UV and FIR emission using the three proxies for this `decoupling' described in Section \ref{sec:Processing_the_Data}. 
The results are shown in Figure \ref{fig:Residuals_against_dust}, in which the mean $\Delta$ in dex for each parameter  is plotted against the different measures for the degree of separation. Each of the five panels shows the residuals for one of the properties plotted against the degree of separation\slash correlation as measured by the three proxies. The coloured lines indicate the median residual in log-spaced bins, while the coloured shaded areas show the mean range enclosed by the 16th and 84th percentiles (i.e. the typical $1\sigma$ error in the limit of Gaussian statistics), and the grey shaded area shows the 16th and 84th percentile range of the scatter within each bin. The bin occupancy is shown by the grey background histogram relative to the right-hand axis. In many cases the scatter is larger than the typical uncertainties, this is likely to be the result of two effects. Firstly, it reflects the fact that the \magphys\ results contain a range of uncertainties that cannot be adequately summarized by a single error bar (the uncertainties show significant variation and contain outliers). Secondly, the uncertainties produced by \magphys\ are likely to be underestimates. This is inevitably the case since the range of SEDs contained in any pre-computed library must by definition be smaller than the actual range of galaxy SEDs in the Universe; for example neither real galaxies or those in our simulations have truly parametric SFHs. In addition, the \magphys\ libraries may not be equally appropriate at all stages of our simulations. 

The average performance of \magphys\ is remarkably consistent, both as a function of the peak-to-peak distance between the UV and FIR images, and as a function of the light-weighted mean UV to FIR distance. In these cases, the mean $\Delta$ is less than $\pm\,0.3$ dex for all parameters, across the separations ranging from 0 to 10\,kpc. In the lower plot of each panel we show the corresponding variation in $\Delta$ (in dex) as a function of the Spearman $\rho$ calculated by comparing the UV and FIR images (recall that only the brightest 20 per cent of pixels were included in this calculation). Here again, the logarithmic difference between the derived and true properties appears independent of $\rho$ once the mean uncertainties are taken in to account.

\section{Discussion}

\subsection{The redshift dependence of the \magphys\ fit success rate}
\label{sec:Sampling bias}

In section \ref{sec:Success rate for fitting an SED} we showed that the fit success rate was a strong function of redshift, with 83\,per cent of the mock observations having acceptable $\chi^2$ overall, but no good fits being obtained at $z > 5.9$. Fixing each mock to be observed at $z = 2$ (Run B) resulted in an increase in the overall success rate to 93 per cent. A likely explanation for this is that the number of SFHs in the \magphys\ library is a strong function of redshift (shown as the dashed line in figure \ref{fig:chi2_success_by_view}, due to the requirement of considering only SFHs shorter than the Hubble time at the observed redshift), which results in significantly worse sampling of the priors at early epochs, particularly when the SFHs of galaxies are so weakly constrained by photometry \citep[e.g.][]{Smith2018}. 

\begin{figure}
    \centering
    \includegraphics[width=0.95\columnwidth]{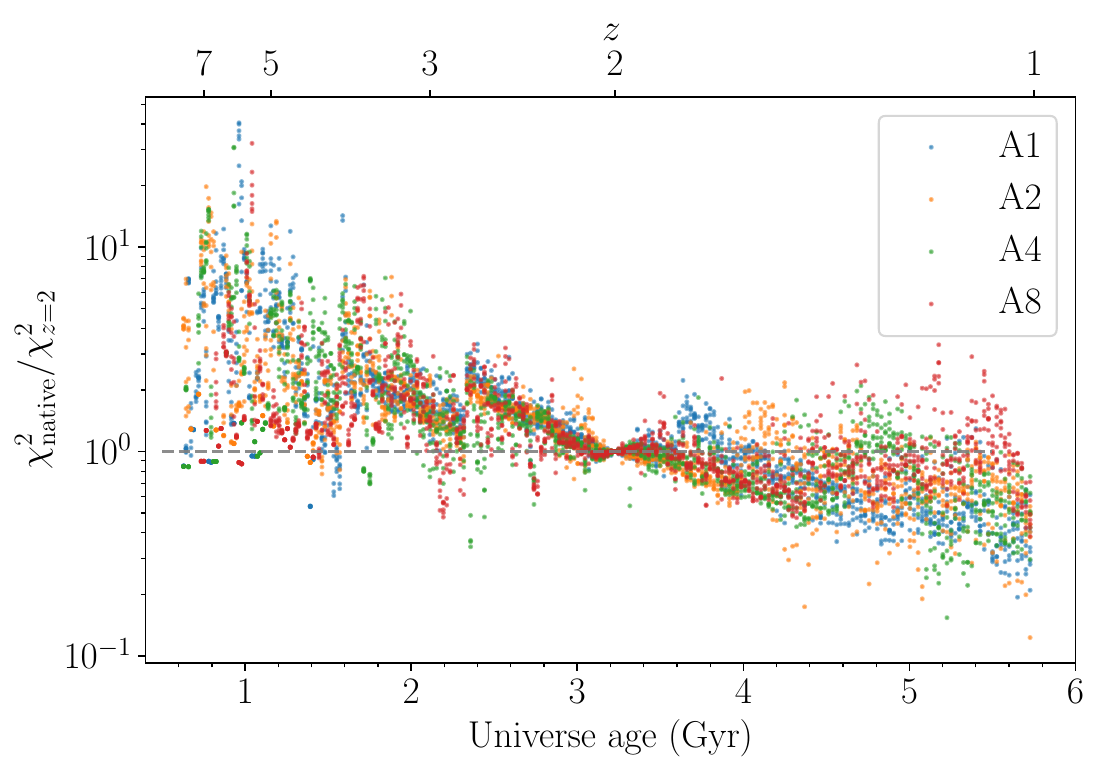}
    \caption{The best-fit $\chi^2$ depends on the size of the \magphys\ library, which varies with the redshift assumed for the fit. This plot shows the ratio of best-fit $\chi^2$ obtained for run A (at the native redshift) to that obtained in run B (where all SEDs were fixed to $z=2$). For galaxies on the left-hand side of this plot the prior gets larger in run B, while for galaxies viewed at later times, the opposite effect is apparent.}
    \label{fig:chi2_ratio}
\end{figure}

In support of this idea, Figure \ref{fig:chi2_ratio} shows the ratio of the best-fit $\chi^2$ obtained for our fiducial results (native redshift run A) to the corresponding value for the SEDs fixed to $z=2$ (run B). It is clear that there is a systematic trend for the native $\chi^2$ to be worse at $z > 2$ (corresponding to a Universe age of $\le 3.2$\,Gyr in our adopted cosmology) and better at $z < 2$. However this trend is by no means absolute, indicating that other effects such as the precise details of the rest-wavelengths being sampled {\color{black} and the number of available filters may also be playing a role.} 

Interestingly, that the ratio of $\chi^2$ for run A to that of run B does not converge on the right-hand side of this plot may indicate that the size of the \magphys\ prior library still impacts the fit quality even at $z < 2$, though of course the difference is that at these comparatively late epochs the priors are sufficiently well-sampled to obtain statistically acceptable fits to the data. 

\subsection{The fidelity of \magphys\ results for dusty, high-redshift galaxies}

The principal aim of this study is to determine how the fidelity of the energy balance code \magphys\ is impacted when it is applied to high-redshift galaxies for which the observed UV and FIR emission are offset, or spatially `decoupled'. For such galaxies, the observed UV light potentially originates from young star clusters that are not spatially co-located with the young stars that dominate the dust heating and thus FIR emission. {{\color{black}Consequently, it is possible that the relatively unobscured young stars could yield a blue UV-optical slope and cause SED modeling codes to underestimate the attenuation. It has been shown that the use of panchromatic data is important when fitting such galaxies \citep{Roebuck2019}, and fitters such as \magphys\ use energy balance to produce physically motivated, panchromatic models that seek to minimise this underestimation. We determine the efficacy of this approach by}
 analyzing the logarithmic difference, $\Delta$, between the true and median-likelihood estimates for stellar mass, SFR, specific SFR, dust mass and dust luminosity as a function of three proxies for the degree of `decoupling' between the UV and FIR data.

In all cases, the performance of \magphys\ appears independent of the degree of UV\slash FIR `decoupling' as measured by all three proxies. We therefore conclude that energy balance SED fitting codes can perform just as well in the presence of such effects as they do when the dust and young stars are co-located within a galaxy. 

We suspect that the explanation for this success is that the \citet{Charlot2000} dust attenuation model used by \magphys\ is sufficiently flexible to handle this `decoupling' in many cases and that the $\chi^2$ algorithm is doing its job by identifying cases for which the model cannot yield a self-consistent solution (i.e. very low attenuation but high FIR luminosity). This has been shown to be the case for an un-modeled AGN contribution to the SED: \citet{Smith2021} noted that using the $\chi^2$ threshold from \citet{Smith2012b}, which we have also implemented here, had the effect of flagging the vast majority of LOFAR-detected AGN as bad fits unless the AGN contribution to the emergent luminosity was very small. {\color{black} Of course, it is expected \citep[e.g.][]{Witt2000} and observed \citep[e.g.][]{Kriek2013,Boquien2022,Nagaraj2022} that the attenuation law is not universal and instead varies by galaxy type. Should additional flexibility be required in future, we note that other works have explored implementing modifications to the standard dust law, including \citet{Battisti2019} who added a $2175\text{\r{A}}$ feature to remove a systematic redshift effect, as well as \citet{LoFaro2017} and \citet{Trayford2020} who allowed the power law indices of equations \ref{eqn:dust1} \&\ \ref{eqn:dust2} to vary. 
However, the fact that there is no scope to easily modify the dust parameterisation assumed in \magphys\ leaves us no option but to defer further investigation of this potentially important aspect for a future work.} 

The reason that some have claimed that energy balance should fail in galaxies with significant IR-UV offsets is that the unobscured lines of sight should dominate the UV emission, meaning that the attenuation that would be inferred from the observed UV-optical emission would be less than the total attenuation experienced by the stellar population as a whole.  However, energy balance codes such as \magphys\ use the FIR luminosity as a simultaneous constraint on the attenuation, and it would simply not be possible to obtain a satisfactory fit to both the UV-optical and FIR regions of the SED assuming low attenuation when the FIR luminosity is high.\footnote{It is tempting to investigate this by making a plot similar to figure \ref{fig:Residuals_against_dust} but including only those fits that exceed the $\chi^2$ threshold we use to identify the bad fits. However, since the best-fit model is statistically unacceptable, we cannot believe the parameter estimates produced by \magphys\ in these cases, meaning that such a test is not meaningful.} Furthermore, we note that even in `normal' galaxies that do not exhibit significant UV-FIR offsets, stars of a given age are not all subject to the same amount of attenuation (e.g. the \citealt{Charlot2000} dust model). Instead, even for a single age and line of sight, there is a distribution of dust optical depths, and this distribution varies with both the stellar age and line of sight considered. The \citet{Charlot2000} model attempts to capture this complex age and line of sight dependence using only two effective optical depths. Though this underlying model is certainly very crude compared to both the simulations and real galaxies, HS15 have already shown that it is adequate to correct for the effects of dust attenuation in at least some low-redshift galaxies. There is no \textit{a priori} reason to believe that it should `break' above some offset threshold (which was the motivation for this study). Our results demonstrate that even when the width of the optical depth distribution experienced by young stars is very wide (i.e. in our simulations some young stars are almost completely unobscured, whereas others have line-of-sight UV optical depths $>>1$), the \citet{Charlot2000} model can still adequately capture the overall effects of dust attenuation in most cases.

\section{Conclusions}

{\color{black}Recent works (e.g. \citealt[][]{Hodge2016,casey2017,Miettinen2017,Simpson2017,Buat2019}) have questioned whether energy balance SED fitting algorithms are appropriate for studying high-redshift star-forming galaxies, due to observations of offsets between the UV and FIR emission \citep[e.g.][]{Hodge2016,Rujopakarn2016,Chen2017,Bowler2018,Gomez2018,Rujopakarn2019}. Clumpy dust distributions within these galaxies may cause a small fraction of relatively unobscured young stars to influence the blue UV-optical
slope and result in an underestimation of the attenuation even if the bulk of the young stars are completely dust-obscured.
We have used four cosmological zoom-in simulations of dusty, high-redshift galaxies from the FIRE-2 project, together with the radiative transfer code SKIRT, to generate over 6,700 synthetic galaxy SEDs spanning a redshift range $8>z>1$. We used these model data to test the fidelity of the galaxy properties recovered using the energy balance fitting code \magphys\ with 18 bands of UV--FIR photometry, building on our previous related studies (HS15, \citealt[][]{Smith2015,Smith2018}).} Our principal findings are as follows: 

\begin{itemize}
  \item We find that the high-$z$ version of \magphys\ was able to produce statistically acceptable best-fit SEDs for 83\,per cent of the synthetic SEDs that we trialled. The fit success rate fell to 50\,per cent for galaxies at $z > 4.85$ and zero for galaxies at $z>5.9$. This reduction in fit success rate has two main contributing factors:
  \begin{enumerate}
      \item the fixed \magphys\ libraries, combined with the requirement that model SFHs should be shorter than the age of the Universe at any given redshift reduces the size of the \magphys\ library available at higher redshifts, mean that the priors become increasingly poorly sampled at earlier times; 
      \item the evolution of the simulated galaxies is increasingly stochastic at the earliest times in our simulations due to their lower mass, causing bursts of star formation to have a disproportionate influence on a galaxy's bolometric luminosity that cannot be reconciled with the \magphys\ prior libraries.
  \end{enumerate}
      
  \item Where statistically acceptable best-fits were obtained, we found that \magphys\ fits are able to broadly capture the true evolution of the four zoom-in simulations that we studied (steady build-up of stellar mass, generally decreasing sSFR, evolution of dust mass), despite individual snapshots being fit independently. In addition, we find that the fidelity of this recovery is remarkably consistent across a broad range of galaxy properties sampled by the simulations, showing no evidence for strong systematics as a function of stellar mass, SFR, sSFR, dust mass or dust luminosity. 
    
  \item Combining UV to FIR observations with an energy balance SED fitting code provides a powerful way to combine multi-wavelength data, and obtain the most reliable estimates of the ground-truth galaxy properties. {\color{black}The panchromatic results outperform those obtained by using either the stellar or dust emission alone.  }  
  
  \item  We find no evidence that the performance of \magphys\ depends on the degree of spatial `decoupling' between the UV and FIR data, despite suggestions to the contrary by several other works. Indeed, our results show that the fidelity of the galaxy properties derived is very similar to that observed for local galaxies, e.g. in our previous work \citep{Hayward2015}. 
\end{itemize}

\section*{Acknowledgements}

{\color{black}The authors thank the anonymous referee for their careful review of this paper.} PH \&\ DJBS would like to thank the Flatiron Institute for their hospitality. The Flatiron Institute is supported by the Simons Foundation. PH would also like to thank Dr Alyssa Drake for helpful input and suggestions. DJBS acknowledges support from the UK Science and Technology Facilities Council (STFC) under grant ST/V000624/1. This research has made use of NASA's Astrophysics Data System Bibliographic Services. DAA acknowledges support by NSF grants AST-2009687 and AST-2108944, CXO grant TM2-23006X, and Simons Foundation award CCA-1018464.

\section*{Data Availability}
The data presented in this paper will be shared on reasonable request to the first author.
Selected snapshots of the four simulated galaxies analyzed here are available as part of the FIRE-2 public data release, see \citet{Wetzel2022} and http://flathub.flatironinstitute.org/fire



\bibliographystyle{mnras}
\bibliography{library} 




\appendix

\section{MAGPHYS fidelity across the range of property values}
\label{first appendix}
As noted in the text, we analysed the systematic recovery of galaxy parameters in more detail by looking at the $\Delta$ for each parameter as a function of the true values of every other parameter. We do not detect any significant trends once the typical error bars (calculated as the mean of the difference between the 16th and 84th percentiles and the median likelihood value) and scatter on the derived values are accounted for. The fidelity of the \magphys\ parameter recovery  persists across a wide range of parameter space.

\begin{figure*}
\centering
\vspace{-1 cm}
\begin{subfigure}{0.9\textwidth}
\includegraphics[width=\textwidth]{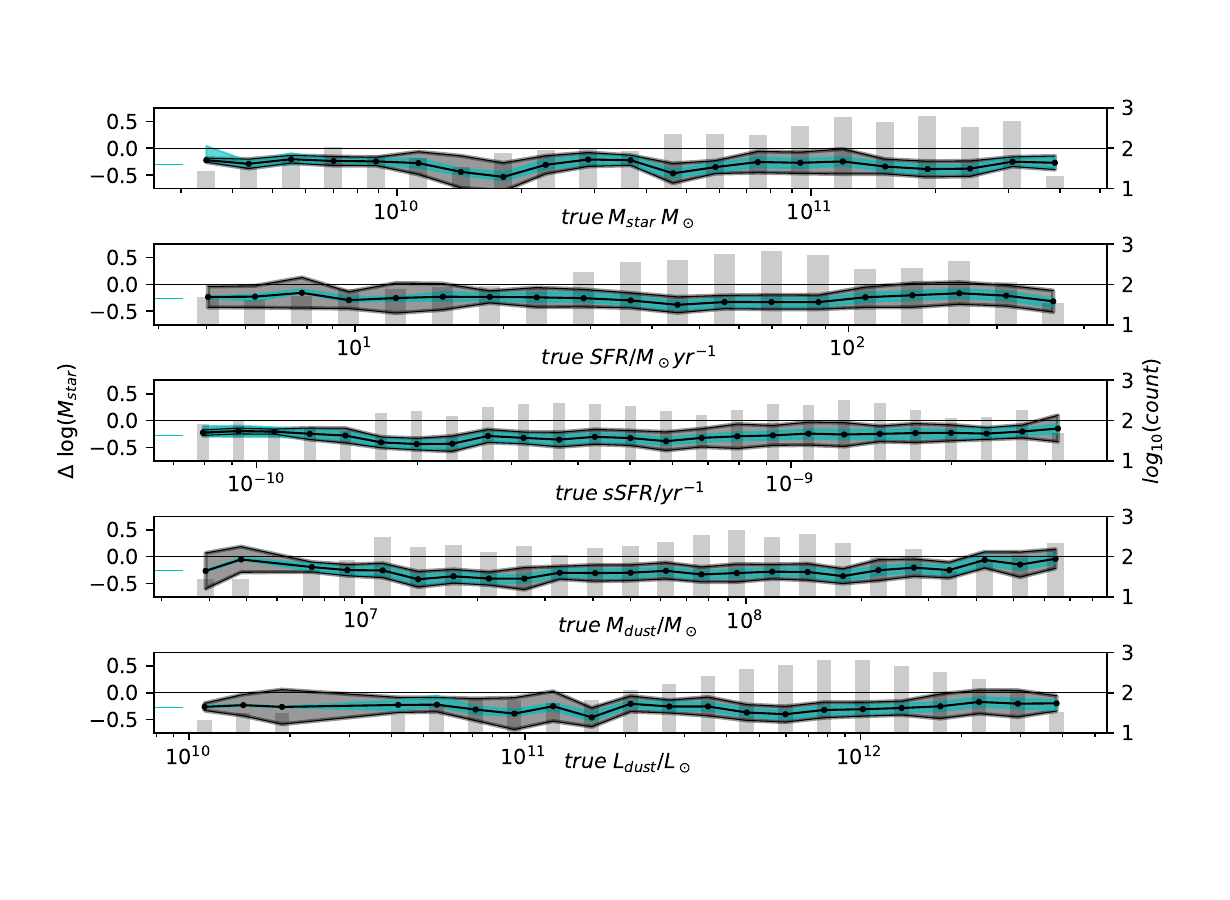}
\vspace{-2 cm}
\caption{$M_{\mathrm{star}}$}
\end{subfigure}
\vspace{-0.5 cm}
\begin{subfigure}{0.9\textwidth}
\includegraphics[width=\textwidth]{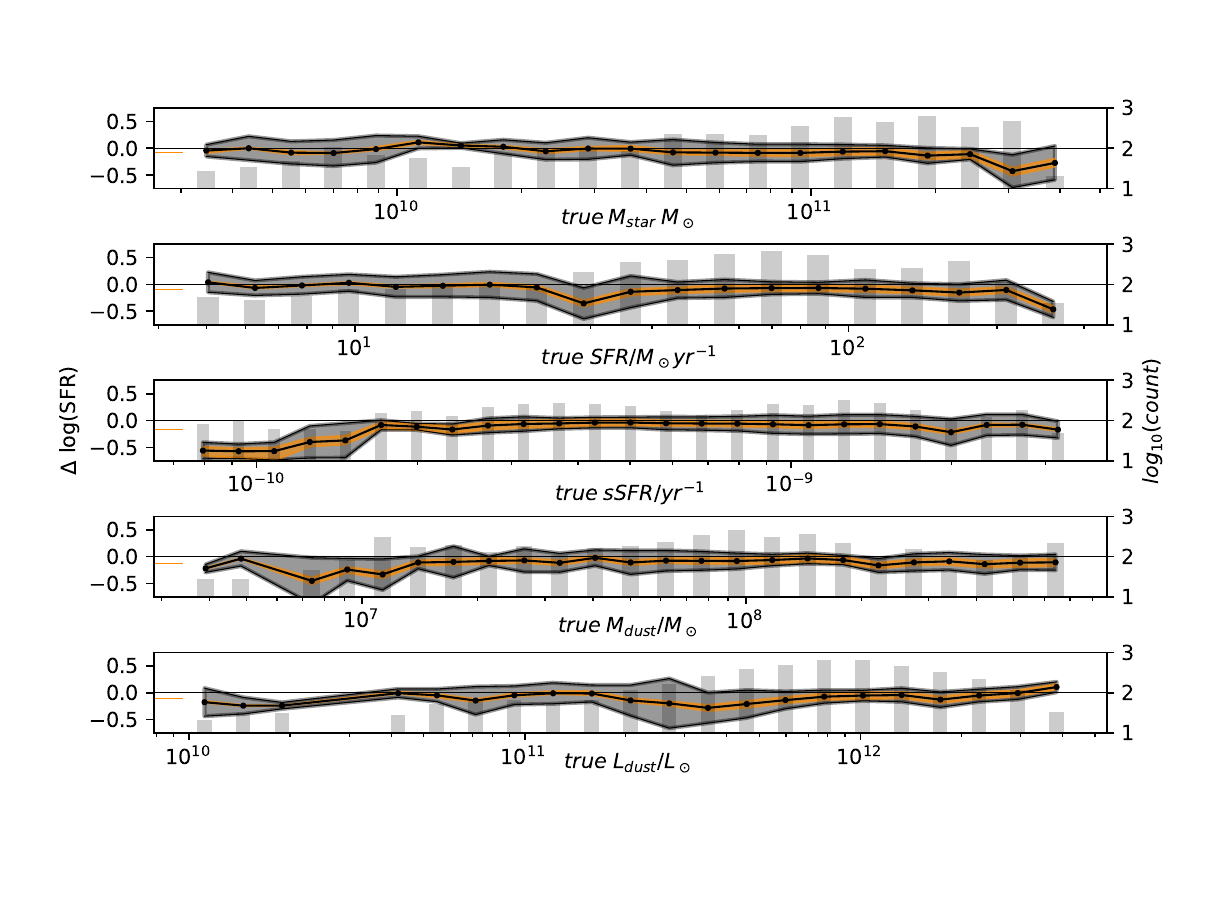}
\vspace{-2 cm}
\caption{SFR}
\end{subfigure}
\vspace{1 cm}
\caption{The fidelity of \magphys' recovery of $M_{\mathrm{star}}$ and SFR is remarkably consistent across the full range of true galaxy properties. The top five panels plot the relationship between the $M_{\mathrm{star}}$ residual and true value of the properties $M_{\mathrm{star}}$, SFR, sSFR, $M_{\mathrm{dust}}$, and $L_{\mathrm{dust}}$ respectively. The data points in black represent the median value for the residual in log-spaced bins; bin occupancy is shown by the background grey bar chart with log values read from the right-hand axis - note that bins with occupancy < 20 have been removed for clarity. In each case the coloured band shows the median 16th and 84th percentile limits for the residuals within the bin and the bounded grey region shows the median 16th and 84th percentile limits for the scatter within the bin. The short coloured line on left-hand of each plot shows the average for the plotted value, residual values are read from the left-hand axis. 
The lower five panels show the same for the SFR residual. }
\label{fig:mstar_sfr_fidelity}
\end{figure*}

\begin{figure*}
\centering
\vspace{-1 cm}
\begin{subfigure}{0.9\textwidth}
\includegraphics[width=\textwidth]{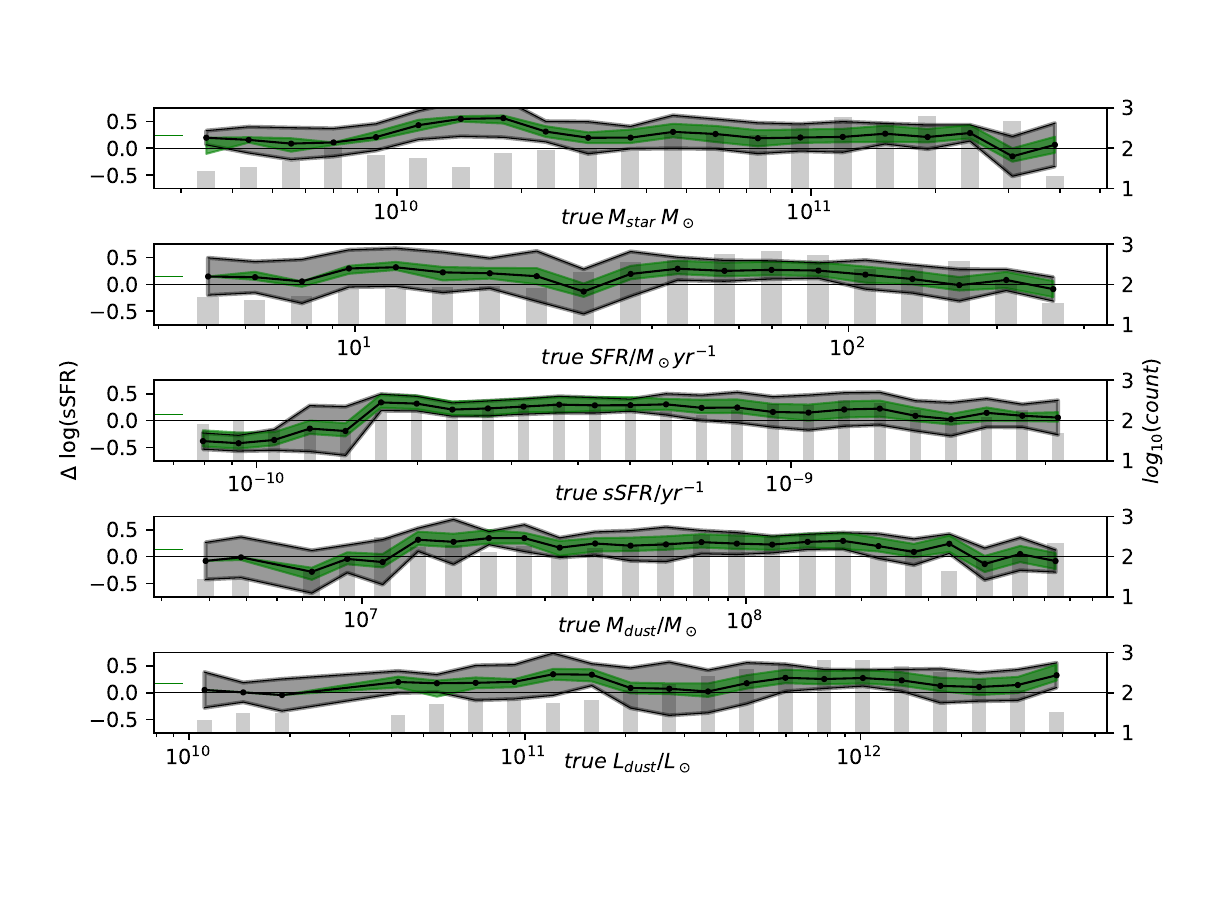}
\vspace{-2 cm}
\caption{sSFR}
\end{subfigure}
\vspace{-0.5 cm}
\begin{subfigure}{0.9\textwidth}
\includegraphics[width=\textwidth]{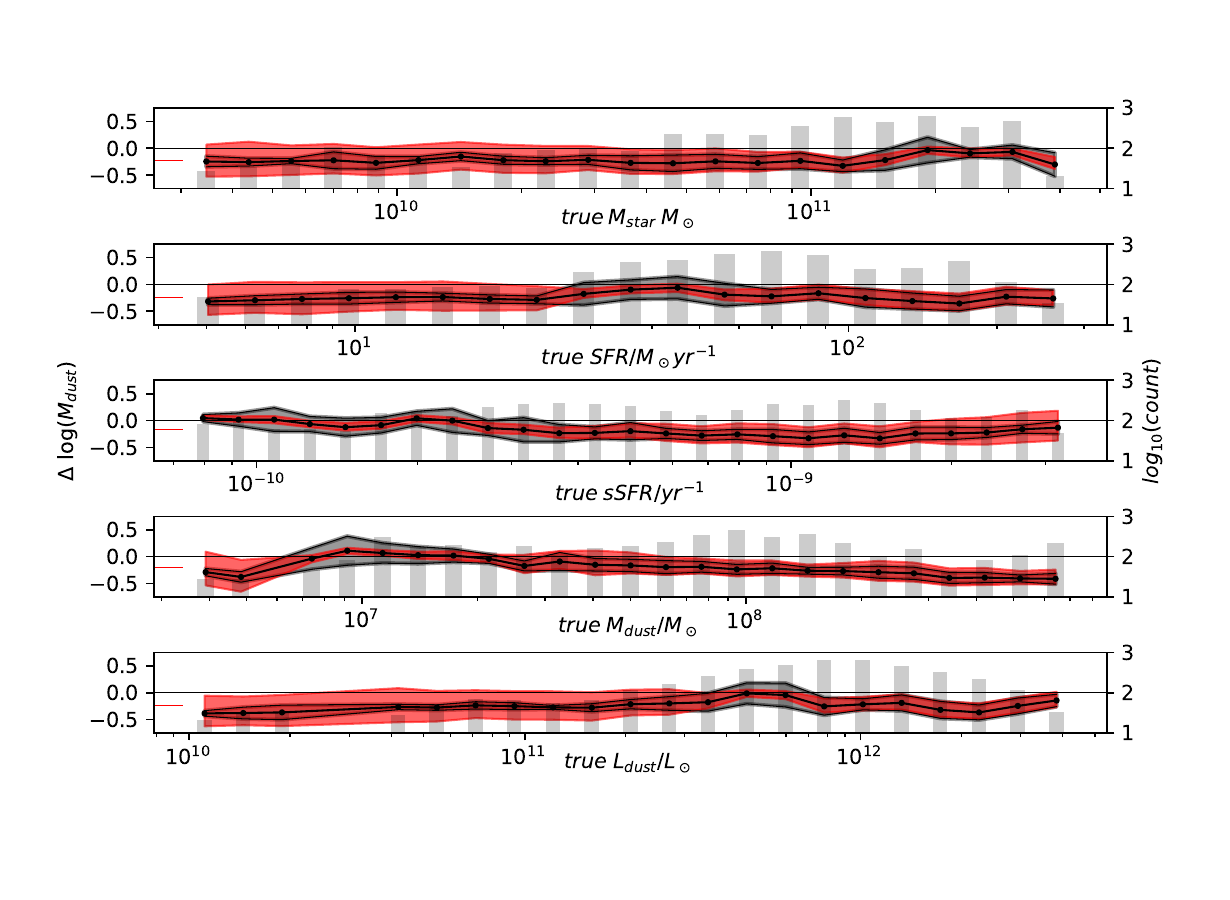}
\vspace{-2 cm}
\caption{M$_\mathrm{dust}$}
\end{subfigure}
\vspace{1 cm}
\caption{Similar to figure \ref{fig:mstar_sfr_fidelity}, but showing the remarkably consistent recovery of sSFR and $M_{\mathrm{dust}}$ as a function of the true galaxy properties. The top five panels plot the relationship between the sSFR residual and true value of the properties $M_{\mathrm{star}}$, SFR, sSFR, $M_{\mathrm{dust}}$, and $L_{\mathrm{dust}}$ respectively. The data points in black represent the median value for the residual in log-spaced bins; bin occupancy is shown by the background grey bar chart with log values read from the right-hand axis - note that bins with occupancy < 20 have been removed for clarity. In each case the coloured band shows the median 16th and 84th percentile limits for the residuals within the bin and the bounded grey region shows the median 16th and 84th percentile limits for the scatter within the bin. The short coloured line on left-hand of each plot shows the average for the plotted value, residual values are read from the left-hand axis. 
The lower five panels show the same for the $M_{\mathrm{dust}}$ residual.}
\end{figure*}

\begin{figure*}
\centering
\vspace{-1 cm}
\includegraphics[width=0.9\textwidth]{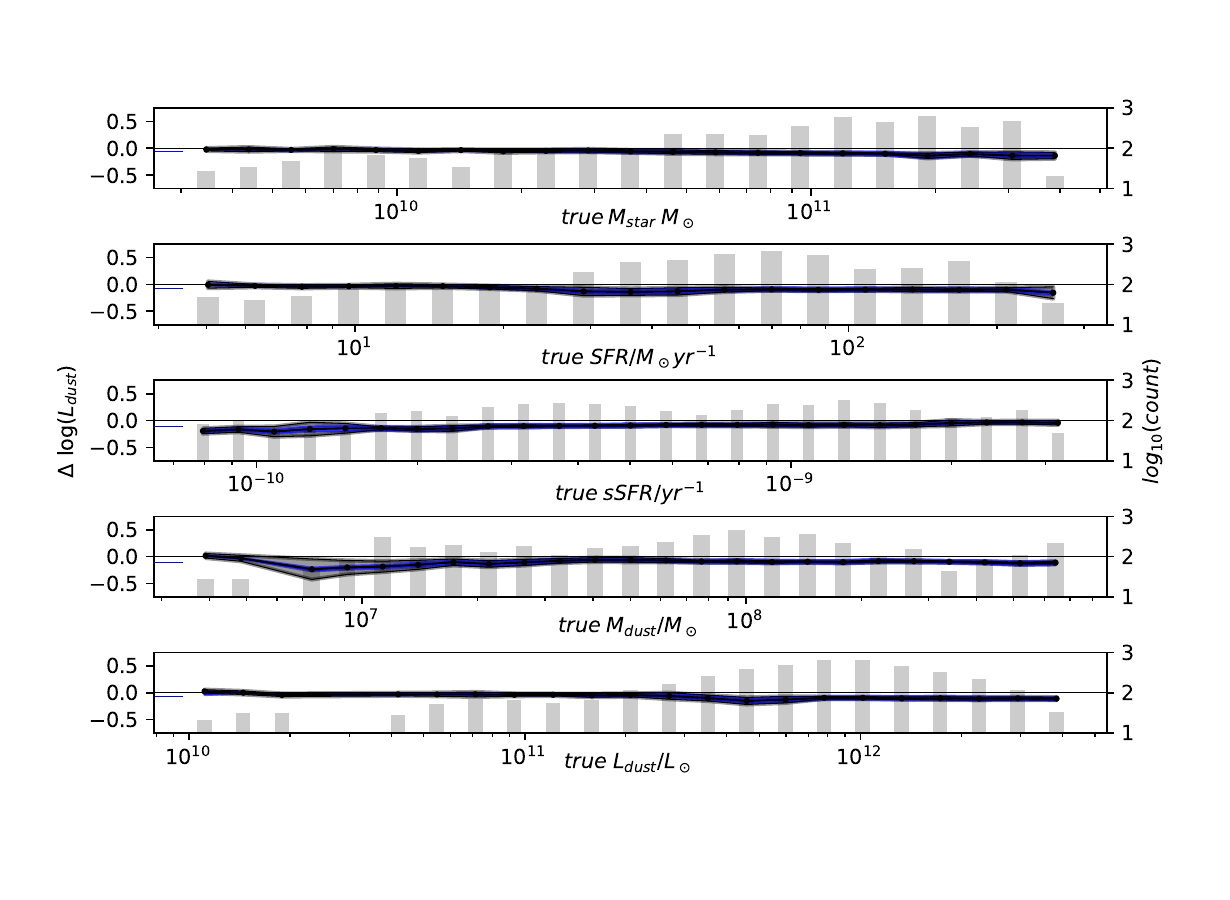}
\vspace{-1 cm}
\caption{Similar to figure \ref{fig:mstar_sfr_fidelity}, but showing the remarkably consistent recovery of $L_{\mathrm{dust}}$. The five panels plot the relationship between the $L_{\mathrm{dust}}$ residual and true value of the properties $M_{\mathrm{star}}$, SFR, sSFR, $M_{\mathrm{dust}}$, and $L_{\mathrm{dust}}$ respectively. The data points in black represent the median value for the residual in log-spaced bins; bin occupancy is shown by the background grey bar chart with log values read from the right-hand axis - note that bins with occupancy < 20 have been removed for clarity. In each case the coloured band shows the median 16th and 84th percentile limits for the residuals within the bin and the bounded grey region shows the median 16th and 84th percentile limits for the scatter within the bin. The short coloured line on left-hand of each plot shows the average for the plotted value, residual values are read from the left hand axis. }
\end{figure*}


\bsp	
\label{lastpage}
\end{document}